\documentclass[%
reprint,
 amsmath,amssymb,
 aps,twocolumn
]{revtex4-1}

\usepackage{float}
\usepackage{graphicx}
\usepackage{dcolumn}
\usepackage{bm}
\usepackage[flushleft]{threeparttable}
\usepackage{url}
\usepackage{amsmath}
\usepackage{amssymb}
\usepackage{braket}
\usepackage{multirow}
\usepackage{array}
\usepackage{url}
\newcolumntype{P}[1]{>{\centering\arraybackslash}p{#1}}
\usepackage{xcolor}
\usepackage{framed}

\colorlet{shadecolor}{blue!80}
\makeatletter
\newcommand*{\rom}[1]{\expandafter\@slowromancap\romannumeral #1@}
\def\yq#1{\textcolor{black}{#1}}

\begin{document}


\preprint{APS/123-QED}

\title{Electron-lattice coupling effects in polarization  switching of charge-order-induced ferroelectrics}

\author{Yubo Qi, and Karin M. Rabe}

\affiliation{%
Department of Physics $\&$ Astronomy, Rutgers University, \\
Piscataway, New Jersey 08854, United States
}%

\pacs{Valid PACS appear here}

\begin{abstract}

We carry out first-principles density-functional-theory calculations to elucidate the polarization switching mechanism in charge-ordering-induced ferroelectrics based on the prototypical case of the (SrVO$_3$)$_1$(LaVO$_3$)$_1$ superlattice.
\yq{We find that lattice relaxation for a specific charge ordering state can ``lock'' that state in, making non-adiabatic switching to a different CO variant energetically prohibitive, and in some cases, even making the energy barrier for adiabatic switching prohibitively large. }
We classify charge-ordering materials into two types, polyhedral breathing and off-centering displacement, based on the type of lattice mode most strongly coupled to the charge ordering.
We demonstrate that the non-adiabatic electron hopping induced by an external electric field is expected only in off-centering-displacement-type charge-ordering-induced ferroelectrics. This successfully explains the different observed switching behaviors of LuFe$_2$O$_4$ and Fe$_3$O$_4$.
These results offer a new understanding of the polarization switching mechanism in charge-ordering-induced ferroelectrics that provides guidance for the design and discovery of charge-ordering-induced ferroelectric materials and \yq{suggests a strategy for realizing ``electronic ferroelectricity'' with polarization switching on electronic rather than lattice time scales. }

\end{abstract}

\maketitle

In strongly correlated materials, the charges in transition metal ions can disproportionate and form an ordered arrangement.
This charge ordering (CO) breaks symmetries and may induce ferroelectricity~\cite{Imada98p1039,Efremov04p853,Van08p434217,Kobayashi12p237601}; this behavior has been experimentally observed in Fe$_3$O$_4$~\cite{Alexe09p4452,Yamauchi09p212404}.
The search for additional CO-induced ferroelectric materials has attracted lively interest. 
Proposed systems include  Pr$_x$Ca$_{1-x}$MnO$_3$~\cite{Goodenough55p564,Daoud02p097205,Efremov05p1433,Efremov04p853,Mercone04p174433,Jardon99p475,Wollan1955neutron,Grenier04p134419,Wu07p174210}, rare earth (R) nickelates (RNiO$_3$)~\cite{Alonso99p3871,Mizokawa00p11263} and rare earth manganites RMn$_2$O$_5$~\cite{Harris07p054447,Kagomiya03p167,Hur04p392,Betouras07p257602,Chapon04p177402,Blake05p214402,Wang07p177202,Lim18p045115}.

Theoretical design based on first-principles calculations has proved valuable in the discovery of CO-induced ferroelectrics~\cite{Park17p087602,Park19p23972}.
The fundamental design principle is that each unit call should possess at least
two multiple-valence atoms, with at least one ordering arrangement that can break the symmetry forbidding polarization.
Our previous first-principles calculations demonstrated that the 1:1 superlattice of SrVO$_3$ and LaVO$_3$ [denoted by (SrVO$_3$)$_1$(LaVO$_3$)$_1$] is a very simple example that satisfies these criteria~\cite{Park17p087602}.
Further first-principles calculations showed that La$_{1/3}$Sr$_{2/3}$FeO$_3$ is another simple example, in which the Fe ions disproportionate into Fe$^{3+}$ and Fe$^{5+}$~\cite{Park19p23972,Krick16p1500372}.

\yq{
In CO-induced ferroelectrics, polarization switching can in principle be produced by non-adiabatic inter-ionic electron hopping alone. CO-induced ferroelectrics are thus natural candidates for electronic ferroelectric materials, which have recently attracted considerable interest for ultrafast switching applications since electron hopping can occur on time scales much shorter than the lattice-driven switching in conventional ferroelectrics~\cite{Ishihara10p011010,Yamauchi14p103201,Alexe09p4452}.
The challenge in realization of electronic ferroelectricity is that the symmetry breaking of the CO state will always be coupled to corresponding distortions of the lattice.
The lattice relaxation for a given CO state generally will favor that state relative to the oppositely-polarized CO state, so that the oppositely-polarized CO state may become too high in energy for the non-adiabatic inter-ionic electron hopping to this state to be driven by an applied electric field. 
In fact, the lattice relaxation energy could be so large that the state gets locked in, and even adiabatic switching is impossible.
Therefore, it is essential to determine the character and strength of the coupling of the CO state to the lattice in the process of design and discovery of CO-induced ferroelectrics and in the realization of electronic ferroelectricity. 
}

In this work, we carry out first-principles calculations on the (SrVO$_3$)$_1$(LaVO$_3$)$_1$ superlattice to investigate the coupling between the CO state and the lattice, describing lattice distortions in terms of symmetry-adapted lattice modes (see supplementary materials (SM) section \rom{1} for computational details~\cite{SM}).
We identify two main types of relevant lattice modes, polyhedral breathing and off-centering displacement, and demonstrate that different types of lattice modes have different degrees of influence on the energy of a CO state.
Specifically, we find that non-adiabatic electron transfer can occur for systems in which the primary mode in lattice relaxation for a given CO state is off-centering displacement, but not for polyhedral-breathing dominated relaxations.
We apply this model to two other representative CO materials, Fe$_3$O$_4$ and LuFe$_2$O$_4$, and show that it successfully explains why Fe$_3$O$_4$ is observed to switch with a small switching polarization while LuFe$_2$O$_4$ is not observed to switch under applied electric field.
This work deepens our understanding of the polarization switching mechanism in CO-induced ferroelectrics and provides valuable guidance for designing CO-induced ferroelectrics and realizing electronic ferroelectricity with polarization switching on electronic time scales.

\begin{figure}[hpbt]
\includegraphics[width=8.5cm]{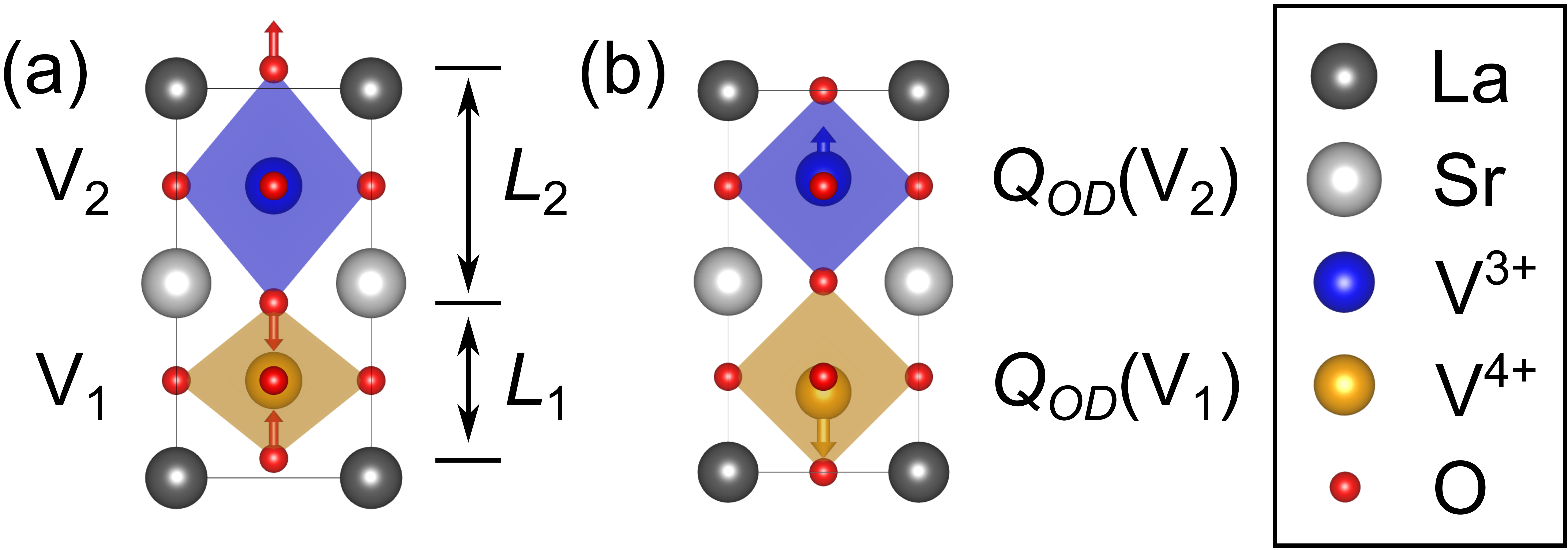}
\caption{Schematic plots of the (a) PB and (b) OD lattice modes in the (SrVO$_3$)$_1$(LaVO$_3$)$_1$ superlattice.}
\label{f1}
\end{figure}

\yq{In previous theoretical work, it was shown that at low temperatures, the (SrVO$_3$)$_1$(LaVO$_3$)$_1$ superlattice is a Mott insulator with vanadium ions disproportionating into V$^{3+}$ and V$^{4+}$~\cite{Park17p087602}.
This superlattice can adopt three different CO patterns which have quite similar energies when the structure is fully relaxed~\cite{Park17p087602}.
Specifically, there is a layered charge ordering (LCO) pattern that can be stabilized over the other two competing patterns by strain or an applied electric field (see SM section \rom{4}. A for the structure);
this pattern, combining with the symmetry breaking by the cation order in the superlattice, breaks symmetry to a polar state.}
Since the ionic radius of V$^{3+}$ is larger than that of V$^{4+}$, we expect a polyhedral breathing (PB) lattice distortion, \yq{as is characteristic of CO materials~\cite{Balachandran13p054101,Mizokawa00p11263,Mazin07p176406}}. \yq{Since the superlattice stacking constrains the in-plane lattice parameter to be equal in the V$^{3+}$ and the V$^{4+}$ layers,} the oxygen octahedron surrounding a V$^{3+}$ ion elongates in the direction perpendicular to the layers and the one surrounding a V$^{4+}$ ion shortens  [Fig.~\ref{f1} (a)].
The PB amplitude can be measured by the ratio of the heights of the two octahedra as 
\begin{equation}
R=\frac{L_1}{L_2}.
\end{equation}
Conversely, if we fix the value of $R$, the V$^{4+}$ ion will tend to favor the site with the smaller value of $L$.


\yq{B-cation off centering displacement (OD, whose magnitude is denoted by $Q_{OD}$) is another lattice distortion that is expected to couple to the ionic size difference in B-site cation charge ordering, with the smaller ion showing a greater tendency to displace.} This is seen in the relaxed structure of the (SrVO$_3$)$_1$(LaVO$_3$)$_1$ superlattice [Fig.~\ref{f1} (b)].
The displacements of V$_1$ and V$_2$ are in opposite directions, pointing from the SrO layer to the LaO layer (see SM section \rom{2} for more discussions).
Since the displacements of the two V ions are much smaller than their interatomic distance, the contribution to polarization from displacements is much smaller than that from the charge ordering pattern.
\yq{As we will discuss further below, we find that the coupling of the off-centering displacement to the charge order is smaller than the coupling of the polyhedral breathing. However, if the polyhedral breathing is suppressed, for example by constraints arising in particular crystal structures, (such as in Fe$_3$O$_4$ discussed later), the off-centering displacement can be the primary lattice mode coupling to the CO state.}

\yq{In contrast, the octahedral rotation distortions ubiquitous in perovskite oxides do not couple strongly to the ionic size difference in B cation ordering. For the (SrVO$_3$)$_1$(LaVO$_3$)$_1$ superlattice, this was established in Ref.~\cite{Park17p087602}, and the octahedral rotation distortions are not discussed further in the present work.}

We perform first-principles density-functional-theory (DFT) calculations on the (SrVO$_3$)$_1$(LaVO$_3$)$_1$ superlattice. We find that the mode magnitudes of the optimized (SrVO$_3$)$_1$(LaVO$_3$)$_1$ superlattice structure are $R_0=0.91$, $Q_{OD}({\rm{V}}^{3+})=0.04$ \AA\ and $Q_{OD}({\rm{V}}^{4+})=0.09$ \AA, consistent with the larger size of $\rm{V}^{3+}$ as discussed above.
The polarization is 0.37 C/m$^2$.
Moreover, we would like to emphasize that it is the arrangement of ${\rm{V}}^{3+}$ and ${\rm{V}}^{4+}$ (the CO state) that primarily determines the polarization (${\rm{V}}_1,{\rm{V}}_2={\rm{V}}^{3+},{\rm{V}}^{4+}$ for the up-polarized state or ${\rm{V}}_1,{\rm{V}}_2={\rm{V}}^{4+},{\rm{V}}^{3+}$ for the down-polarized state).
The OD can influence the stability of a particular CO state (as we will discuss later), but such displacements themselves have much less contribution to the polarization, compared with the contribution of the CO.

\begin{figure}[hpbt]
\includegraphics[width=8.5cm]{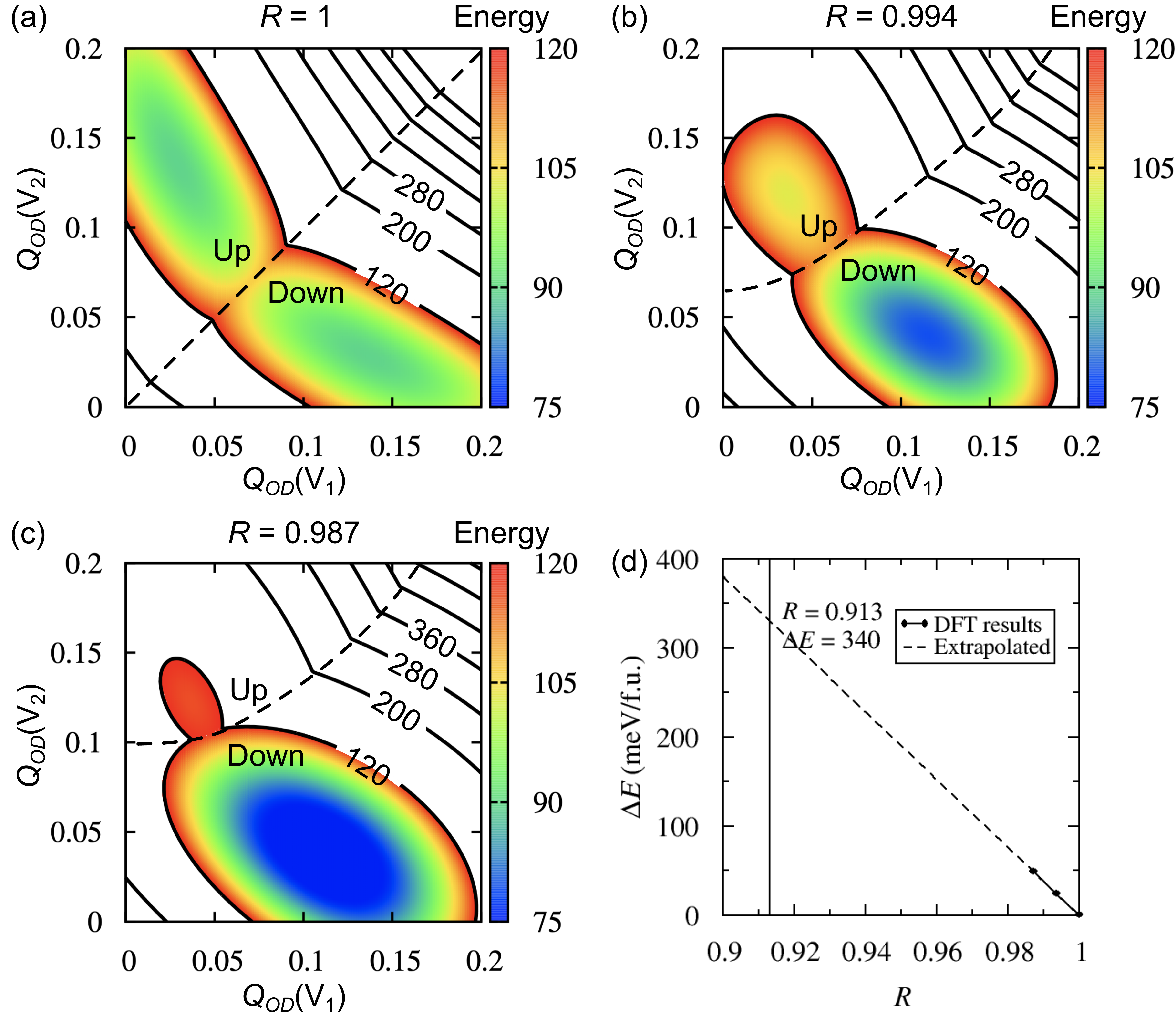}
\caption{Energy landscapes as functions of $Q_{OD}({\rm{V}}_{1})$ and $Q_{OD}({\rm{V}}_{2})$, 
with (a) $R=1$, (b) $R=0.994$, and (c) $R=0.987$. The dashed line represents the boundary between the regions corresponding to the up- and down-polarized states. (d) $R$ {\em{vs.}} $\Delta{E}$ with available data,
based on which we estimate that $\Delta{E}=340$ meV/f.u. at $R=R_0=0.913$.}
\label{f2}
\end{figure}

To investigate the coupling between the lattice modes and CO states, 
we generate energy landscapes as described in SM section \rom{1} for various $R$ values in the plane of $Q_{OD}({\rm{V}}_{1})$ and $Q_{OD}({\rm{V}}_{2})$~\cite{Han18p067601,Georgescu19p14434,Georgescu2021energy}.
We use occupation matrix control to generate starting electronic states with ${\rm{V}}_1,{\rm{V}}_2={\rm{V}}^{3+},{\rm{V}}^{4+}$ and ${\rm{V}}_1,{\rm{V}}_2={\rm{V}}^{4+},{\rm{V}}^{3+}$ respectively, relax the structures with fixed PB and OD modes, and select the lowest-energy CO state for each structure.
\yq{We begin with the energy landscape for $R=1$. 
Electronic correlation promotes localization of the electron into a single ionic orbital (see SM section \rom{3} for the occupation matrices).
Even with polyhedral breathing completely suppressed and $Q_{OD}({\rm{V}}_{1})=Q_{OD}({\rm{V}}_{2})$, we find that charge disproportionation is still energetically favorable.
With $Q_{OD}({\rm{V}}_{1})\ {\neq}\ Q_{OD}({\rm{V}}_{2})$, the preferred arrangement of the V$^{3+}$ and V$^{4+}$ will be determined by the OD modes.}
The resulting energy landscape is shown in Fig.~\ref{f2} (a). 
In the upper part, we have 
\begin{equation}
Q_{OD}({\rm{V}}_{1})<Q_{OD}({\rm{V}}_{2})\Rightarrow{\rm{V}}_1={\rm{V}}^{3+}\ {\rm{and}}\ {\rm{V}}_2={\rm{V}}^{4+},
\end{equation}
which corresponds to an up-polarized CO state.
Similarly, the lower part corresponds to a down-polarized CO state. 
Because of the symmetry imposed by $R=1$,
the energies of the two local minima corresponding to up- and down-polarized states are identical (with symmetry-related structures), and the difference $\Delta{E}=0$.
As $R$ decreases to 0.994 [Fig.~\ref{f2} (b)], the oxygen octahedron associated with ${\rm{V}}_{1}$ compresses, favoring  ${\rm{V}}^{4+}$.
The effect on the energy landscape is that the dashed line, which represents the boundary of the up- and down-polarized regions, moves toward the left corner, indicating that $Q_{OD}({\rm{V}}_{2})$ needs to exceed $Q_{OD}({\rm{V}}_{1})$ by a critical amount to stabilize the up-polarized state.
The energy difference between the two local minima $\Delta{E}=E_{\rm{up}}-E_{\rm{down}}$ becomes 24 meV/f.u..
As $R$ decreases to 0.987 [Fig.~\ref{f2} (c)], $\Delta{E}$ increases to 49 meV/f.u., and the local minimum corresponding to the up-polarized CO state becomes quite shallow.
If $R$ decreases further to 0.981 (not shown in Fig.~\ref{f2}), the up-polarized CO state completely loses its stability; that is, no matter how
much bigger $Q_{OD}({\rm{V}}_{2})$ is than $Q_{OD}({\rm{V}}_{1})$, ${\rm{V}}^{3+}$ will favor the ${\rm{V}}_{2}$ site.
\yq{This is consistent with the well-known phenomenon of stabilization of a particular CO state by lattice distortion, with PB modes being the most strongly coupled and leading to the greatest energy gains relative to other distortions such as OD.}

The $R$ value at which the second local minimum disappears is still far from the R value for the relaxed structure $R_0=0.913$.
Since there is no local minimum corresponding to the up-polarized CO state for $R=R_0$,
we cannot calculate $\Delta{E}$ directly. 
To get an estimate, we plot the computed results of $R$ {\em{vs.}} $\Delta{E}$ in Fig.~\ref{f2} (d), do a linear extrapolation, and infer that $\Delta{E}=340$ meV/f.u. at $R=0.913$.
To induce an electron transfer between ${\rm{V}}_{1}$ and ${\rm{V}}_{2}$, 
this energy difference has to be compensated by the electrical enthalpy 
\begin{equation}
\Delta{E}=\Delta{P}{\cdot}E_{el},
\end{equation}
where $\Delta{P}$ is the difference between the polarization of the two CO states, and $E_{el}$ is the applied electric field. 
For $\Delta{E}=340$ meV/f.u., we have
\begin{equation}
E_{el}=\Delta{E}/\Delta{P}=7\ {\rm{MV/cm}}.
\end{equation}
which is larger than the breakdown field of most perovskites~\cite{Shende01p1648,Ueda64p1267}.
\yq{We emphasize that it is the polyhedral breathing relaxation that makes the oppositely-polarized CO state so much higher in energy, making it difficult or impossible to achieve purely electronic polarization switching with accessible applied electric fields. 
}

To investigate the polarization switchability of the (SrVO$_3$)$_1$(LaVO$_3$)$_1$ superlattice, 
we relax the down-polarized structure under a nonzero up-directed electric field, starting from zero field and gradually increasing the field magnitude. At each step, we take the resulting structure, reverse the CO state using the occupation matrix method (SM section \rom{1}), and relax to locate the up-polarized local minimum, if it exists.
We find that the electronically up-polarized state is unstable
for electric fields all the way to 12.6 MV/cm, the largest field for which we could obtain converged results. 
This is as expected, since our analysis above showed that for $R_0=0.913$, the electric field for balancing the two local minima is 7 MV/cm, and the electric field required to overcome the energy barrier should be even larger.
For the relaxed down-polarized structure, the change in the $R$ value with increasing electric field was negligible. 
This is reasonable because the PB mode is infrared inactive and thus does not couple directly to an electric field.

The analysis above suggests that the polarization in a PB-type CO material will be difficult to switch with an applied electric field, because the polarization switching has to be accompanied by a change of the PB lattice mode, which does not couple directly to an electric field. 
However, the situation is different for an OD-type CO material.
To investigate the influence of the OD mode upon the stability of a CO state, we consider the (SrVO$_3$)$_1$(LaVO$_3$)$_1$ superlattice with $R$ artificially fixed to 1.
The PB character is thus completely eliminated, and in zero applied field, the CO state, and thus the polarization, couples to the OD modes only. 
We begin with a down-polarized structure, in which the $4+$ valence state is on the ${\rm{V}}_1$ site.
\yq{To explore whether the non-adiabatic electron transfer can occur in such a artificially constrained OD type CO material,
we consider the corresponding `electronically' up-polarized CO state, which means that its structure is the same as the down-polarized one, but the arrangement of the V$^{3+}$ and V$^{4+}$ is artificially flipped by manipulating the occupation matrices. 
At zero field, the energy of this electronically up-polarized state should be higher than the down-polarized state, since the structure is optimized for the down-polarized CO state.}
We find that the energy difference between the down-polarized and electronically up-polarized states is 109 meV/f.u..
This can be balanced by an electric field above 1.95 MV/cm, a value readily accessible in the laboratory.
This threshold electric field is much reduced compared with that in a PB-type CO material, consistent with weaker coupling of the OD displacement to the CO state. 
It is also worth mentioning that if the structure is allowed to relax with $R$ fixed to 1, the energy difference can be reduced due to the direct coupling of the OD distortion to the electric field, but this effect is secondary, here lowering the threshold electric field for polarization switching from 1.95 MV/cm to 1.62 MV/cm.

\yq{Our results for the (SrVO$_3$)$_1$(LaVO$_3$)$_1$ superlattice structure suggest that PB- and OD-type CO materials will have different electrical behaviors.
For the OD type, the coupling of the distortion to the CO state is relatively weak, so that the opposite electronically-polarized state is low enough in energy to allow non-adiabatic electron hopping driven by applied electric field.
In contrast, the PB lattice mode strongly couples to the CO state, prohibiting non-adiabatic electron hopping to the oppositely-polarized state, and possibly even making the barrier for conventional lattice-based switching prohibitively high. Moreover, this PB mode is infrared inactive and changes little in an electric field. 
This offers an understanding of why polarization switching (let alone electron-hopping time scale switching) has not been observed in some proposed CO-induced ``electronic ferroelectrics'', and what is required to achieve ultrafast switching.
}
In the following, we apply this model to other CO materials, such as LuFe$_2$O$_4$ and Fe$_3$O$_4$, and discuss the implications for the difference in their experimentally-observed electrical behaviors.

First, we discuss polarization switchability in LuFe$_2$O$_4$. In 2005, Ikeda {\em{et al.}} demonstrated that LuFe$_2$O$_4$ has two oppositely polarized states,
which can be obtained by cooling from a high-temperature centrosymmetric structure under oppositely directed electric fields~\cite{Ikeda05p1136,Ikeda08p434218,Isobe90p1917}.
The emergence of polarization was attributed to the disproportionation and ordered arrangement of Fe atoms~\cite{Ikeda05p1136,Ikeda08p434218,Xiang07p246403,Xiang09p132408}.
These results made LuFe$_2$O$_4$ a promising candidate for
CO-induced ferroelectricity. 
However, electric-field-induced polarization switching in LuFe$_2$O$_4$ has not been experimentally demonstrated,
leading to a long-running debate about whether LuFe$_2$O$_4$ is indeed ferroelectric~\cite{de2012p187601,Lafuerza13p085130,Ruff12p290}. 

\begin{figure}[hpbt]
\includegraphics[width=8.5cm]{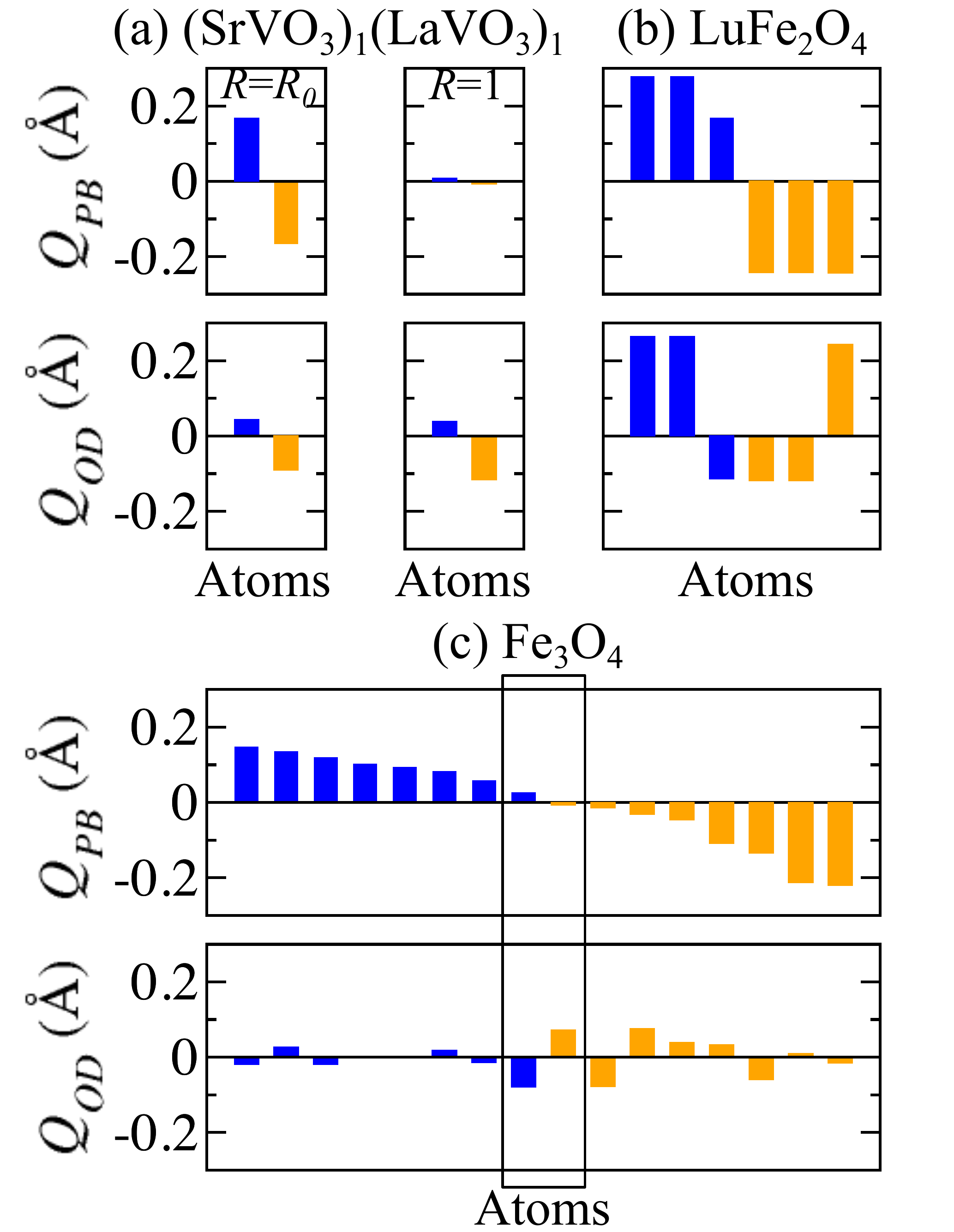}
\caption {$Q_{PB}$ and $Q_{OD}$ of the ions in 
(a) unconstrained ($R=R_0$) and constrained ($R=1$) (SrVO$_3$)$_1$(LaVO$_3$)$_1$ superlattice, and (b) LuFe$_2$O$_4$,
(c) Fe$_3$O$_4$. 
\yq{Each bar represents a transition metal ion in the primitive cell.
The primitive cells of the (SrVO$_3$)$_1$(LaVO$_3$)$_1$ superlattice, LuFe$_2$O$_4$, and Fe$_3$O$_4$ have 2, 6, and 16 non-equivalent transition metal ions respectively.  
The blue and orange bars correspond to the lower valence (V$^{3+}$ or Fe$^{2+}$) and higher valence states (V$^{4+}$ or Fe$^{3+}$) respectively.
The ions are ordered according to their values of $Q_{PB}$.
The box in (c) shows a pair of Fe sites whose valence states can switch.}
} 
\label{f4}
\end{figure}

In the following, we perform an analysis of the previously reported structure~\cite{Xiang09p132408} that demonstrates that LuFe$_2$O$_4$ is a PB-type CO material, whose polarization is difficult to switch with an electric field (see SM section \rom{4}. B for the structure). 
To quantify the polyhedral breathing distortion, for each transition-metal-centered polyhedra, we define $Q_{PB}$ as
\begin{equation}
Q_{PB}(i)=\sum_j\left(r_{ij}-\bar{r}\right),
\end{equation}
where $j$ runs over all the oxygen atoms bonded to the Fe ion $i$, $r_{ij}$ is the distance from the center of the polyhedron containing the Fe ion $i$ to the oxygen atom $j$, 
and $\bar{r}$ is the average of all the $r_{ij}$ in the entire unit cell.
A negative $Q_{PB}$ corresponds to a smaller polyhedron, and a positive $Q_{PB}$ corresponds to a larger polyhedron.
\yq{For the (SrVO$_3$)$_1$(LaVO$_3$)$_1$ superlattice, $R<1$ corresponds to $Q_{PB}\left({\rm{V}}_1\right)<0$ and $Q_{PB}\left({\rm{V}}_2\right)>0$, as shown in Fig.~\ref{f4} (a).}
In Fig.~\ref{f4} (b), we also plot $Q_{PB}$ and $Q_{OD}$ for the six Fe ions in a LuFe$_2$O$_4$ primitive cell~\cite{Xiang09p132408}.
We can see that in LuFe$_2$O$_4$, $Q_{PB}$ for Fe$^{2+}$ is noticeably larger than $Q_{PB}$ for Fe$^{3+}$, with a ratio even larger than that of the fully relaxed (SrVO$_3$)$_1$(LaVO$_3$)$_1$ superlattice [Fig.~\ref{f4} (a) $R=R_0$]. 
From this we conclude that the CO states in LuFe$_2$O$_4$ are strongly coupled to the PB mode.

Our previous analysis has demonstrated that the polarization in PB-type CO materials generally cannot be switched by an external electric field.
Therefore, we conclude that the CO-induced polarization in LuFe$_2$O$_4$ is unswitchable under an applied electric field.
However,
each CO state can be obtained by annealing the high-symmetry R$\bar{3}$m structure from high temperatures down to $T<T_{\rm{CO}}$ with a bias field~\cite{Ikeda05p1136}, consistent with experimental observation.

Next, we consider magnetite (Fe$_3$O$_4$), which is known to be a CO-induced ferroelectric with switchable polarization~\cite{Alexe09p4452,Yamauchi13p113703,Yamauchi09p212404,Senn12p173}.
Below the Verwey temperature $T_{\rm{V}}=125$ K, Fe$_3$O$_4$ adopts a monoclinic $Cc$ structure~\cite{Verwey39p327}.
Previous first-principles calculations have located the Fe$^{2+}$ and Fe$^{3+}$ at the 16 inequivalent 6-fold sites of this structure and shown that they are non-centrosymmetrically arranged~\cite{Yamauchi09p212404}, leading to a non-zero polarization (see SM section \rom{4}. C for the structure).
In Fig.~\ref{f4} (c), we plot the $Q_{PB}$ and $Q_{OD}$ of these 16 Fe ions, in descending order of their $Q_{PB}$. We note that the 8 Fe$^{2+}$ ions have positive $Q_{PB}$ and 8 Fe$^{3+}$ ions have negative $Q_{PB}$~\cite{Yamauchi09p212404}, which further supports that the PB mode has a strong influence on the stability of CO states.
We now focus on the pair of ions in the middle of the range in Fig.~\ref{f4} (c), one Fe$^{2+}$ and one Fe$^{3+}$, which we note are neighboring in the crystal structure (see SM section \rom{4}. C). 
They have quite similar $Q_{PB}$ and noticeably different $Q_{OD}$ with opposite signs, analogous to the constrained (SrVO$_3$)$_1$(LaVO$_3$)$_1$ superlattice with $R=1$.
According to our analysis above, electric-field-induced electron hopping can occur from the Fe$^{2+}$ to the Fe$^{3+}$.
The change of polarization due to the valence switching of this pair of Fe ions is 3 $\mu$C/m$^2$,
which is the same order of magnitude as the experimental result (5 $\mu$C/m$^2$ in Ref.~\cite{Ziese12p086007}).

\yq{In summary, we propose a polarization switching model in CO materials based on first-principles calculations and lattice mode analysis.  
Specifically, we demonstrate that whether a non-adiabatic inter-ion electron transfer can occur depends on the type of lattice distortion, polyhedral breathing or off-centering displacement mode, which primarily couples to the CO state.}
Based on our first-principles analysis of the \yq{theoretically designed ferroelectric} (SrVO$_3$)$_1$(LaVO$_3$)$_1$ superlattice,
we demonstrate that electron-transfer-induced polarization switching can only occur in a system that has at least a subset of sites for which the strongest coupling is to the off-centering displacements rather than to the polyhedral breathing.
Our theory successfully explains the experimentally-observed different switchability of LuFe$_2$O$_4$ and Fe$_3$O$_4$.
This work presents a new understanding of the polarization switching mechanism in charge-ordering-induced ferroelectrics.
\yq{It explains why the polarization in some proposed charge-ordering-induced ``electronic ferroelectrics'' cannot in fact be switched by an electric field, and provides valuable insights and strategies for the design and discovery of electronic ferroelectrics with polarization switching on electronic time scales.}

\section*{acknowledgement}
This work was supported by ONR N00014-21-1-2107. 
First-principles calculations were performed using the computational resources provided by the Rutgers University Parallel Computing (RUPC) clusters and the High-Performance Computing Modernization Office of the Department of Defense.

\bibliography{cite3}

\begin{thebibliography}{62}%
\makeatletter
\providecommand \@ifxundefined [1]{%
 \@ifx{#1\undefined}
}%
\providecommand \@ifnum [1]{%
 \ifnum #1\expandafter \@firstoftwo
 \else \expandafter \@secondoftwo
 \fi
}%
\providecommand \@ifx [1]{%
 \ifx #1\expandafter \@firstoftwo
 \else \expandafter \@secondoftwo
 \fi
}%
\providecommand \natexlab [1]{#1}%
\providecommand \enquote  [1]{``#1''}%
\providecommand \bibnamefont  [1]{#1}%
\providecommand \bibfnamefont [1]{#1}%
\providecommand \citenamefont [1]{#1}%
\providecommand \href@noop [0]{\@secondoftwo}%
\providecommand \href [0]{\begingroup \@sanitize@url \@href}%
\providecommand \@href[1]{\@@startlink{#1}\@@href}%
\providecommand \@@href[1]{\endgroup#1\@@endlink}%
\providecommand \@sanitize@url [0]{\catcode `\\12\catcode `\$12\catcode
  `\&12\catcode `\#12\catcode `\^12\catcode `\_12\catcode `\%12\relax}%
\providecommand \@@startlink[1]{}%
\providecommand \@@endlink[0]{}%
\providecommand \url  [0]{\begingroup\@sanitize@url \@url }%
\providecommand \@url [1]{\endgroup\@href {#1}{\urlprefix }}%
\providecommand \urlprefix  [0]{URL }%
\providecommand \Eprint [0]{\href }%
\providecommand \doibase [0]{http://dx.doi.org/}%
\providecommand \selectlanguage [0]{\@gobble}%
\providecommand \bibinfo  [0]{\@secondoftwo}%
\providecommand \bibfield  [0]{\@secondoftwo}%
\providecommand \translation [1]{[#1]}%
\providecommand \BibitemOpen [0]{}%
\providecommand \bibitemStop [0]{}%
\providecommand \bibitemNoStop [0]{.\EOS\space}%
\providecommand \EOS [0]{\spacefactor3000\relax}%
\providecommand \BibitemShut  [1]{\csname bibitem#1\endcsname}%
\let\auto@bib@innerbib\@empty
\bibitem [{\citenamefont {Imada}\ \emph {et~al.}(1998)\citenamefont {Imada},
  \citenamefont {Fujimori},\ and\ \citenamefont {Tokura}}]{Imada98p1039}%
  \BibitemOpen
  \bibfield  {author} {\bibinfo {author} {\bibfnamefont {M.}~\bibnamefont
  {Imada}}, \bibinfo {author} {\bibfnamefont {A.}~\bibnamefont {Fujimori}}, \
  and\ \bibinfo {author} {\bibfnamefont {Y.}~\bibnamefont {Tokura}},\
  }\href@noop {} {\bibfield  {journal} {\bibinfo  {journal} {Rev. Mod. Phys.}\
  }\textbf {\bibinfo {volume} {70}},\ \bibinfo {pages} {1039} (\bibinfo {year}
  {1998})}\BibitemShut {NoStop}%
\bibitem [{\citenamefont {Efremov}\ \emph {et~al.}(2004)\citenamefont
  {Efremov}, \citenamefont {Van Den~Brink},\ and\ \citenamefont
  {Khomskii}}]{Efremov04p853}%
  \BibitemOpen
  \bibfield  {author} {\bibinfo {author} {\bibfnamefont {D.~V.}\ \bibnamefont
  {Efremov}}, \bibinfo {author} {\bibfnamefont {J.}~\bibnamefont {Van
  Den~Brink}}, \ and\ \bibinfo {author} {\bibfnamefont {D.~I.}\ \bibnamefont
  {Khomskii}},\ }\href@noop {} {\bibfield  {journal} {\bibinfo  {journal} {Nat.
  Mater.}\ }\textbf {\bibinfo {volume} {3}},\ \bibinfo {pages} {853} (\bibinfo
  {year} {2004})}\BibitemShut {NoStop}%
\bibitem [{\citenamefont {Van Den~Brink}\ and\ \citenamefont
  {Khomskii}(2008)}]{Van08p434217}%
  \BibitemOpen
  \bibfield  {author} {\bibinfo {author} {\bibfnamefont {J.}~\bibnamefont {Van
  Den~Brink}}\ and\ \bibinfo {author} {\bibfnamefont {D.~I.}\ \bibnamefont
  {Khomskii}},\ }\href@noop {} {\bibfield  {journal} {\bibinfo  {journal} {J.
  Phys. Condens. Matter}\ }\textbf {\bibinfo {volume} {20}},\ \bibinfo {pages}
  {434217} (\bibinfo {year} {2008})}\BibitemShut {NoStop}%
\bibitem [{\citenamefont {Kobayashi}\ \emph {et~al.}(2012)\citenamefont
  {Kobayashi}, \citenamefont {Horiuchi}, \citenamefont {Kumai}, \citenamefont
  {Kagawa}, \citenamefont {Murakami},\ and\ \citenamefont
  {Tokura}}]{Kobayashi12p237601}%
  \BibitemOpen
  \bibfield  {author} {\bibinfo {author} {\bibfnamefont {K.}~\bibnamefont
  {Kobayashi}}, \bibinfo {author} {\bibfnamefont {S.}~\bibnamefont {Horiuchi}},
  \bibinfo {author} {\bibfnamefont {R.}~\bibnamefont {Kumai}}, \bibinfo
  {author} {\bibfnamefont {F.}~\bibnamefont {Kagawa}}, \bibinfo {author}
  {\bibfnamefont {Y.}~\bibnamefont {Murakami}}, \ and\ \bibinfo {author}
  {\bibfnamefont {Y.}~\bibnamefont {Tokura}},\ }\href@noop {} {\bibfield
  {journal} {\bibinfo  {journal} {Phys. Rev. Lett.}\ }\textbf {\bibinfo
  {volume} {108}},\ \bibinfo {pages} {237601} (\bibinfo {year}
  {2012})}\BibitemShut {NoStop}%
\bibitem [{\citenamefont {Alexe}\ \emph {et~al.}(2009)\citenamefont {Alexe},
  \citenamefont {Ziese}, \citenamefont {Hesse}, \citenamefont {Esquinazi},
  \citenamefont {Yamauchi}, \citenamefont {Fukushima}, \citenamefont
  {Picozzi},\ and\ \citenamefont {G{\"o}sele}}]{Alexe09p4452}%
  \BibitemOpen
  \bibfield  {author} {\bibinfo {author} {\bibfnamefont {M.}~\bibnamefont
  {Alexe}}, \bibinfo {author} {\bibfnamefont {M.}~\bibnamefont {Ziese}},
  \bibinfo {author} {\bibfnamefont {D.}~\bibnamefont {Hesse}}, \bibinfo
  {author} {\bibfnamefont {P.}~\bibnamefont {Esquinazi}}, \bibinfo {author}
  {\bibfnamefont {K.}~\bibnamefont {Yamauchi}}, \bibinfo {author}
  {\bibfnamefont {T.}~\bibnamefont {Fukushima}}, \bibinfo {author}
  {\bibfnamefont {S.}~\bibnamefont {Picozzi}}, \ and\ \bibinfo {author}
  {\bibfnamefont {U.}~\bibnamefont {G{\"o}sele}},\ }\href@noop {} {\bibfield
  {journal} {\bibinfo  {journal} {Adv. Mater.}\ }\textbf {\bibinfo {volume}
  {21}},\ \bibinfo {pages} {4452} (\bibinfo {year} {2009})}\BibitemShut
  {NoStop}%
\bibitem [{\citenamefont {Yamauchi}\ \emph {et~al.}(2009)\citenamefont
  {Yamauchi}, \citenamefont {Fukushima},\ and\ \citenamefont
  {Picozzi}}]{Yamauchi09p212404}%
  \BibitemOpen
  \bibfield  {author} {\bibinfo {author} {\bibfnamefont {K.}~\bibnamefont
  {Yamauchi}}, \bibinfo {author} {\bibfnamefont {T.}~\bibnamefont {Fukushima}},
  \ and\ \bibinfo {author} {\bibfnamefont {S.}~\bibnamefont {Picozzi}},\
  }\href@noop {} {\bibfield  {journal} {\bibinfo  {journal} {Phys. Rev. B}\
  }\textbf {\bibinfo {volume} {79}},\ \bibinfo {pages} {212404} (\bibinfo
  {year} {2009})}\BibitemShut {NoStop}%
\bibitem [{\citenamefont {Goodenough}(1955)}]{Goodenough55p564}%
  \BibitemOpen
  \bibfield  {author} {\bibinfo {author} {\bibfnamefont {J.~B.}\ \bibnamefont
  {Goodenough}},\ }\href@noop {} {\bibfield  {journal} {\bibinfo  {journal}
  {Phys. Rev.}\ }\textbf {\bibinfo {volume} {100}},\ \bibinfo {pages} {564}
  (\bibinfo {year} {1955})}\BibitemShut {NoStop}%
\bibitem [{\citenamefont {Daoud-Aladine}\ \emph {et~al.}(2002)\citenamefont
  {Daoud-Aladine}, \citenamefont {Rodriguez-Carvajal}, \citenamefont
  {Pinsard-Gaudart}, \citenamefont {Fernandez-Diaz},\ and\ \citenamefont
  {Revcolevschi}}]{Daoud02p097205}%
  \BibitemOpen
  \bibfield  {author} {\bibinfo {author} {\bibfnamefont {A.}~\bibnamefont
  {Daoud-Aladine}}, \bibinfo {author} {\bibfnamefont {J.}~\bibnamefont
  {Rodriguez-Carvajal}}, \bibinfo {author} {\bibfnamefont {L.}~\bibnamefont
  {Pinsard-Gaudart}}, \bibinfo {author} {\bibfnamefont {M.}~\bibnamefont
  {Fernandez-Diaz}}, \ and\ \bibinfo {author} {\bibfnamefont {A.}~\bibnamefont
  {Revcolevschi}},\ }\href@noop {} {\bibfield  {journal} {\bibinfo  {journal}
  {Phys. Rev. Lett.}\ }\textbf {\bibinfo {volume} {89}},\ \bibinfo {pages}
  {097205} (\bibinfo {year} {2002})}\BibitemShut {NoStop}%
\bibitem [{\citenamefont {Efremov}\ \emph {et~al.}(2005)\citenamefont
  {Efremov}, \citenamefont {van~den Brink},\ and\ \citenamefont
  {Khomskii}}]{Efremov05p1433}%
  \BibitemOpen
  \bibfield  {author} {\bibinfo {author} {\bibfnamefont {D.~V.}\ \bibnamefont
  {Efremov}}, \bibinfo {author} {\bibfnamefont {J.}~\bibnamefont {van~den
  Brink}}, \ and\ \bibinfo {author} {\bibfnamefont {D.~I.}\ \bibnamefont
  {Khomskii}},\ }\href@noop {} {\bibfield  {journal} {\bibinfo  {journal}
  {Physica B}\ }\textbf {\bibinfo {volume} {359}},\ \bibinfo {pages} {1433}
  (\bibinfo {year} {2005})}\BibitemShut {NoStop}%
\bibitem [{\citenamefont {Mercone}\ \emph {et~al.}(2004)\citenamefont
  {Mercone}, \citenamefont {Wahl}, \citenamefont {Pautrat}, \citenamefont
  {Pollet},\ and\ \citenamefont {Simon}}]{Mercone04p174433}%
  \BibitemOpen
  \bibfield  {author} {\bibinfo {author} {\bibfnamefont {S.}~\bibnamefont
  {Mercone}}, \bibinfo {author} {\bibfnamefont {A.}~\bibnamefont {Wahl}},
  \bibinfo {author} {\bibfnamefont {A.}~\bibnamefont {Pautrat}}, \bibinfo
  {author} {\bibfnamefont {M.}~\bibnamefont {Pollet}}, \ and\ \bibinfo {author}
  {\bibfnamefont {C.}~\bibnamefont {Simon}},\ }\href@noop {} {\bibfield
  {journal} {\bibinfo  {journal} {Phys. Rev. B}\ }\textbf {\bibinfo {volume}
  {69}},\ \bibinfo {pages} {174433} (\bibinfo {year} {2004})}\BibitemShut
  {NoStop}%
\bibitem [{\citenamefont {Jard{\'o}n}\ \emph {et~al.}(1999)\citenamefont
  {Jard{\'o}n}, \citenamefont {Rivadulla}, \citenamefont {Hueso}, \citenamefont
  {Fondado}, \citenamefont {Lopez-Quintela}, \citenamefont {Rivas},
  \citenamefont {Zysler}, \citenamefont {Causa},\ and\ \citenamefont
  {Sanchez}}]{Jardon99p475}%
  \BibitemOpen
  \bibfield  {author} {\bibinfo {author} {\bibfnamefont {C.}~\bibnamefont
  {Jard{\'o}n}}, \bibinfo {author} {\bibfnamefont {F.}~\bibnamefont
  {Rivadulla}}, \bibinfo {author} {\bibfnamefont {L.}~\bibnamefont {Hueso}},
  \bibinfo {author} {\bibfnamefont {A.}~\bibnamefont {Fondado}}, \bibinfo
  {author} {\bibfnamefont {M.~A.}\ \bibnamefont {Lopez-Quintela}}, \bibinfo
  {author} {\bibfnamefont {J.}~\bibnamefont {Rivas}}, \bibinfo {author}
  {\bibfnamefont {R.}~\bibnamefont {Zysler}}, \bibinfo {author} {\bibfnamefont
  {M.~T.}\ \bibnamefont {Causa}}, \ and\ \bibinfo {author} {\bibfnamefont
  {R.~D.}\ \bibnamefont {Sanchez}},\ }\href@noop {} {\bibfield  {journal}
  {\bibinfo  {journal} {J. Magn. Magn. Mater.}\ }\textbf {\bibinfo {volume}
  {196}},\ \bibinfo {pages} {475} (\bibinfo {year} {1999})}\BibitemShut
  {NoStop}%
\bibitem [{\citenamefont {Wollan}\ and\ \citenamefont
  {Koehler}(1955)}]{Wollan1955neutron}%
  \BibitemOpen
  \bibfield  {author} {\bibinfo {author} {\bibfnamefont {E.~O.}\ \bibnamefont
  {Wollan}}\ and\ \bibinfo {author} {\bibfnamefont {W.~C.}\ \bibnamefont
  {Koehler}},\ }\href@noop {} {\bibfield  {journal} {\bibinfo  {journal} {Phys.
  Rev.}\ }\textbf {\bibinfo {volume} {100}},\ \bibinfo {pages} {545} (\bibinfo
  {year} {1955})}\BibitemShut {NoStop}%
\bibitem [{\citenamefont {Grenier}\ \emph {et~al.}(2004)\citenamefont
  {Grenier}, \citenamefont {Hill}, \citenamefont {Gibbs}, \citenamefont
  {Thomas}, \citenamefont {Zimmermann}, \citenamefont {Nelson}, \citenamefont
  {Kiryukhin}, \citenamefont {Tokura}, \citenamefont {Tomioka}, \citenamefont
  {Casa}, \citenamefont {Gog},\ and\ \citenamefont
  {Venkataraman}}]{Grenier04p134419}%
  \BibitemOpen
  \bibfield  {author} {\bibinfo {author} {\bibfnamefont {S.}~\bibnamefont
  {Grenier}}, \bibinfo {author} {\bibfnamefont {J.~P.}\ \bibnamefont {Hill}},
  \bibinfo {author} {\bibfnamefont {D.}~\bibnamefont {Gibbs}}, \bibinfo
  {author} {\bibfnamefont {K.~J.}\ \bibnamefont {Thomas}}, \bibinfo {author}
  {\bibfnamefont {M.~v.}\ \bibnamefont {Zimmermann}}, \bibinfo {author}
  {\bibfnamefont {C.~S.}\ \bibnamefont {Nelson}}, \bibinfo {author}
  {\bibfnamefont {V.}~\bibnamefont {Kiryukhin}}, \bibinfo {author}
  {\bibfnamefont {Y.}~\bibnamefont {Tokura}}, \bibinfo {author} {\bibfnamefont
  {Y.}~\bibnamefont {Tomioka}}, \bibinfo {author} {\bibfnamefont
  {D.}~\bibnamefont {Casa}}, \bibinfo {author} {\bibfnamefont {T.}~\bibnamefont
  {Gog}}, \ and\ \bibinfo {author} {\bibfnamefont {C.}~\bibnamefont
  {Venkataraman}},\ }\href@noop {} {\bibfield  {journal} {\bibinfo  {journal}
  {Phys. Rev. B}\ }\textbf {\bibinfo {volume} {69}},\ \bibinfo {pages} {134419}
  (\bibinfo {year} {2004})}\BibitemShut {NoStop}%
\bibitem [{\citenamefont {Wu}\ \emph {et~al.}(2007)\citenamefont {Wu},
  \citenamefont {Klie}, \citenamefont {Zhu},\ and\ \citenamefont
  {Jooss}}]{Wu07p174210}%
  \BibitemOpen
  \bibfield  {author} {\bibinfo {author} {\bibfnamefont {L.}~\bibnamefont
  {Wu}}, \bibinfo {author} {\bibfnamefont {R.~F.}\ \bibnamefont {Klie}},
  \bibinfo {author} {\bibfnamefont {Y.}~\bibnamefont {Zhu}}, \ and\ \bibinfo
  {author} {\bibfnamefont {C.}~\bibnamefont {Jooss}},\ }\href@noop {}
  {\bibfield  {journal} {\bibinfo  {journal} {Phys. Rev. B}\ }\textbf {\bibinfo
  {volume} {76}},\ \bibinfo {pages} {174210} (\bibinfo {year}
  {2007})}\BibitemShut {NoStop}%
\bibitem [{\citenamefont {Alonso}\ \emph {et~al.}(1999)\citenamefont {Alonso},
  \citenamefont {Garc{\'\i}a-Mu{\~n}oz}, \citenamefont
  {Fern{\'a}ndez-D{\'\i}az}, \citenamefont {Aranda}, \citenamefont
  {Mart{\'\i}nez-Lope},\ and\ \citenamefont {Casais}}]{Alonso99p3871}%
  \BibitemOpen
  \bibfield  {author} {\bibinfo {author} {\bibfnamefont {J.~A.}\ \bibnamefont
  {Alonso}}, \bibinfo {author} {\bibfnamefont {J.~L.}\ \bibnamefont
  {Garc{\'\i}a-Mu{\~n}oz}}, \bibinfo {author} {\bibfnamefont {M.~T.}\
  \bibnamefont {Fern{\'a}ndez-D{\'\i}az}}, \bibinfo {author} {\bibfnamefont
  {M.~A.~G.}\ \bibnamefont {Aranda}}, \bibinfo {author} {\bibfnamefont {M.~J.}\
  \bibnamefont {Mart{\'\i}nez-Lope}}, \ and\ \bibinfo {author} {\bibfnamefont
  {M.~T.}\ \bibnamefont {Casais}},\ }\href@noop {} {\bibfield  {journal}
  {\bibinfo  {journal} {Phys. Rev. Lett.}\ }\textbf {\bibinfo {volume} {82}},\
  \bibinfo {pages} {3871} (\bibinfo {year} {1999})}\BibitemShut {NoStop}%
\bibitem [{\citenamefont {Mizokawa}\ \emph {et~al.}(2000)\citenamefont
  {Mizokawa}, \citenamefont {Khomskii},\ and\ \citenamefont
  {Sawatzky}}]{Mizokawa00p11263}%
  \BibitemOpen
  \bibfield  {author} {\bibinfo {author} {\bibfnamefont {T.}~\bibnamefont
  {Mizokawa}}, \bibinfo {author} {\bibfnamefont {D.~I.}\ \bibnamefont
  {Khomskii}}, \ and\ \bibinfo {author} {\bibfnamefont {G.~A.}\ \bibnamefont
  {Sawatzky}},\ }\href@noop {} {\bibfield  {journal} {\bibinfo  {journal}
  {Phys. Rev. B}\ }\textbf {\bibinfo {volume} {61}},\ \bibinfo {pages} {11263}
  (\bibinfo {year} {2000})}\BibitemShut {NoStop}%
\bibitem [{\citenamefont {Harris}(2007)}]{Harris07p054447}%
  \BibitemOpen
  \bibfield  {author} {\bibinfo {author} {\bibfnamefont {A.~B.}\ \bibnamefont
  {Harris}},\ }\href@noop {} {\bibfield  {journal} {\bibinfo  {journal} {Phys.
  Rev. B}\ }\textbf {\bibinfo {volume} {76}},\ \bibinfo {pages} {054447}
  (\bibinfo {year} {2007})}\BibitemShut {NoStop}%
\bibitem [{\citenamefont {Kagomiya}\ \emph {et~al.}(2003)\citenamefont
  {Kagomiya}, \citenamefont {Matsumoto}, \citenamefont {Kohn}, \citenamefont
  {Fukuda}, \citenamefont {Shoubu}, \citenamefont {Kimura}, \citenamefont
  {Noda},\ and\ \citenamefont {Ikeda}}]{Kagomiya03p167}%
  \BibitemOpen
  \bibfield  {author} {\bibinfo {author} {\bibfnamefont {I.}~\bibnamefont
  {Kagomiya}}, \bibinfo {author} {\bibfnamefont {S.}~\bibnamefont {Matsumoto}},
  \bibinfo {author} {\bibfnamefont {K.}~\bibnamefont {Kohn}}, \bibinfo {author}
  {\bibfnamefont {Y.}~\bibnamefont {Fukuda}}, \bibinfo {author} {\bibfnamefont
  {T.}~\bibnamefont {Shoubu}}, \bibinfo {author} {\bibfnamefont
  {H.}~\bibnamefont {Kimura}}, \bibinfo {author} {\bibfnamefont
  {Y.}~\bibnamefont {Noda}}, \ and\ \bibinfo {author} {\bibfnamefont
  {N.}~\bibnamefont {Ikeda}},\ }\href@noop {} {\bibfield  {journal} {\bibinfo
  {journal} {Ferroelectrics}\ }\textbf {\bibinfo {volume} {286}},\ \bibinfo
  {pages} {167} (\bibinfo {year} {2003})}\BibitemShut {NoStop}%
\bibitem [{\citenamefont {Hur}\ \emph {et~al.}(2004)\citenamefont {Hur},
  \citenamefont {Park}, \citenamefont {Sharma}, \citenamefont {Ahn},
  \citenamefont {Guha},\ and\ \citenamefont {Cheong}}]{Hur04p392}%
  \BibitemOpen
  \bibfield  {author} {\bibinfo {author} {\bibfnamefont {N.}~\bibnamefont
  {Hur}}, \bibinfo {author} {\bibfnamefont {S.}~\bibnamefont {Park}}, \bibinfo
  {author} {\bibfnamefont {P.}~\bibnamefont {Sharma}}, \bibinfo {author}
  {\bibfnamefont {J.}~\bibnamefont {Ahn}}, \bibinfo {author} {\bibfnamefont
  {S.}~\bibnamefont {Guha}}, \ and\ \bibinfo {author} {\bibfnamefont {S.-W.}\
  \bibnamefont {Cheong}},\ }\href@noop {} {\bibfield  {journal} {\bibinfo
  {journal} {Nature}\ }\textbf {\bibinfo {volume} {429}},\ \bibinfo {pages}
  {392} (\bibinfo {year} {2004})}\BibitemShut {NoStop}%
\bibitem [{\citenamefont {Betouras}\ \emph {et~al.}(2007)\citenamefont
  {Betouras}, \citenamefont {Giovannetti},\ and\ \citenamefont {van~den
  Brink}}]{Betouras07p257602}%
  \BibitemOpen
  \bibfield  {author} {\bibinfo {author} {\bibfnamefont {J.~J.}\ \bibnamefont
  {Betouras}}, \bibinfo {author} {\bibfnamefont {G.}~\bibnamefont
  {Giovannetti}}, \ and\ \bibinfo {author} {\bibfnamefont {J.}~\bibnamefont
  {van~den Brink}},\ }\href@noop {} {\bibfield  {journal} {\bibinfo  {journal}
  {Phys. Rev. Lett.}\ }\textbf {\bibinfo {volume} {98}},\ \bibinfo {pages}
  {257602} (\bibinfo {year} {2007})}\BibitemShut {NoStop}%
\bibitem [{\citenamefont {Chapon}\ \emph {et~al.}(2004)\citenamefont {Chapon},
  \citenamefont {Blake}, \citenamefont {Gutmann}, \citenamefont {Park},
  \citenamefont {Hur}, \citenamefont {Radaelli},\ and\ \citenamefont
  {Cheong}}]{Chapon04p177402}%
  \BibitemOpen
  \bibfield  {author} {\bibinfo {author} {\bibfnamefont {L.~C.}\ \bibnamefont
  {Chapon}}, \bibinfo {author} {\bibfnamefont {G.~R.}\ \bibnamefont {Blake}},
  \bibinfo {author} {\bibfnamefont {M.~J.}\ \bibnamefont {Gutmann}}, \bibinfo
  {author} {\bibfnamefont {S.}~\bibnamefont {Park}}, \bibinfo {author}
  {\bibfnamefont {N.}~\bibnamefont {Hur}}, \bibinfo {author} {\bibfnamefont
  {P.~G.}\ \bibnamefont {Radaelli}}, \ and\ \bibinfo {author} {\bibfnamefont
  {S.-W.}\ \bibnamefont {Cheong}},\ }\href@noop {} {\bibfield  {journal}
  {\bibinfo  {journal} {Phys. Rev. Lett.}\ }\textbf {\bibinfo {volume} {93}},\
  \bibinfo {pages} {177402} (\bibinfo {year} {2004})}\BibitemShut {NoStop}%
\bibitem [{\citenamefont {Blake}\ \emph {et~al.}(2005)\citenamefont {Blake},
  \citenamefont {Chapon}, \citenamefont {Radaelli}, \citenamefont {Park},
  \citenamefont {Hur}, \citenamefont {Cheong},\ and\ \citenamefont
  {Rodriguez-Carvajal}}]{Blake05p214402}%
  \BibitemOpen
  \bibfield  {author} {\bibinfo {author} {\bibfnamefont {G.~R.}\ \bibnamefont
  {Blake}}, \bibinfo {author} {\bibfnamefont {L.~C.}\ \bibnamefont {Chapon}},
  \bibinfo {author} {\bibfnamefont {P.~G.}\ \bibnamefont {Radaelli}}, \bibinfo
  {author} {\bibfnamefont {S.}~\bibnamefont {Park}}, \bibinfo {author}
  {\bibfnamefont {N.}~\bibnamefont {Hur}}, \bibinfo {author} {\bibfnamefont
  {S.-W.}\ \bibnamefont {Cheong}}, \ and\ \bibinfo {author} {\bibfnamefont
  {J.}~\bibnamefont {Rodriguez-Carvajal}},\ }\href@noop {} {\bibfield
  {journal} {\bibinfo  {journal} {Phys. Rev. B}\ }\textbf {\bibinfo {volume}
  {71}},\ \bibinfo {pages} {214402} (\bibinfo {year} {2005})}\BibitemShut
  {NoStop}%
\bibitem [{\citenamefont {Wang}\ \emph {et~al.}(2007)\citenamefont {Wang},
  \citenamefont {Guo},\ and\ \citenamefont {He}}]{Wang07p177202}%
  \BibitemOpen
  \bibfield  {author} {\bibinfo {author} {\bibfnamefont {C.}~\bibnamefont
  {Wang}}, \bibinfo {author} {\bibfnamefont {G.-C.}\ \bibnamefont {Guo}}, \
  and\ \bibinfo {author} {\bibfnamefont {L.}~\bibnamefont {He}},\ }\href@noop
  {} {\bibfield  {journal} {\bibinfo  {journal} {Phys. Rev. Lett.}\ }\textbf
  {\bibinfo {volume} {99}},\ \bibinfo {pages} {177202} (\bibinfo {year}
  {2007})}\BibitemShut {NoStop}%
\bibitem [{\citenamefont {Lim}\ \emph {et~al.}(2018)\citenamefont {Lim},
  \citenamefont {Saldana-Greco},\ and\ \citenamefont {Rappe}}]{Lim18p045115}%
  \BibitemOpen
  \bibfield  {author} {\bibinfo {author} {\bibfnamefont {J.~S.}\ \bibnamefont
  {Lim}}, \bibinfo {author} {\bibfnamefont {D.}~\bibnamefont {Saldana-Greco}},
  \ and\ \bibinfo {author} {\bibfnamefont {A.~M.}\ \bibnamefont {Rappe}},\
  }\href@noop {} {\bibfield  {journal} {\bibinfo  {journal} {Phys. Rev. B}\
  }\textbf {\bibinfo {volume} {97}},\ \bibinfo {pages} {045115} (\bibinfo
  {year} {2018})}\BibitemShut {NoStop}%
\bibitem [{\citenamefont {Park}\ \emph {et~al.}(2017)\citenamefont {Park},
  \citenamefont {Kumar},\ and\ \citenamefont {Rabe}}]{Park17p087602}%
  \BibitemOpen
  \bibfield  {author} {\bibinfo {author} {\bibfnamefont {S.~Y.}\ \bibnamefont
  {Park}}, \bibinfo {author} {\bibfnamefont {A.}~\bibnamefont {Kumar}}, \ and\
  \bibinfo {author} {\bibfnamefont {K.~M.}\ \bibnamefont {Rabe}},\ }\href@noop
  {} {\bibfield  {journal} {\bibinfo  {journal} {Phys. Rev. Lett.}\ }\textbf
  {\bibinfo {volume} {118}},\ \bibinfo {pages} {087602} (\bibinfo {year}
  {2017})}\BibitemShut {NoStop}%
\bibitem [{\citenamefont {Park}\ \emph {et~al.}(2019)\citenamefont {Park},
  \citenamefont {Rabe},\ and\ \citenamefont {Neaton}}]{Park19p23972}%
  \BibitemOpen
  \bibfield  {author} {\bibinfo {author} {\bibfnamefont {S.~Y.}\ \bibnamefont
  {Park}}, \bibinfo {author} {\bibfnamefont {K.~M.}\ \bibnamefont {Rabe}}, \
  and\ \bibinfo {author} {\bibfnamefont {J.~B.}\ \bibnamefont {Neaton}},\
  }\href@noop {} {\bibfield  {journal} {\bibinfo  {journal} {Proc. Natl. Acad.
  Sci.}\ }\textbf {\bibinfo {volume} {116}},\ \bibinfo {pages} {23972}
  (\bibinfo {year} {2019})}\BibitemShut {NoStop}%
\bibitem [{\citenamefont {Krick}\ \emph {et~al.}(2016)\citenamefont {Krick},
  \citenamefont {Lee}, \citenamefont {Sichel-Tissot}, \citenamefont {Rappe},\
  and\ \citenamefont {May}}]{Krick16p1500372}%
  \BibitemOpen
  \bibfield  {author} {\bibinfo {author} {\bibfnamefont {A.~L.}\ \bibnamefont
  {Krick}}, \bibinfo {author} {\bibfnamefont {C.-W.}\ \bibnamefont {Lee}},
  \bibinfo {author} {\bibfnamefont {R.~J.}\ \bibnamefont {Sichel-Tissot}},
  \bibinfo {author} {\bibfnamefont {A.~M.}\ \bibnamefont {Rappe}}, \ and\
  \bibinfo {author} {\bibfnamefont {S.~J.}\ \bibnamefont {May}},\ }\href@noop
  {} {\bibfield  {journal} {\bibinfo  {journal} {Adv. Electron. Mater.}\
  }\textbf {\bibinfo {volume} {2}},\ \bibinfo {pages} {1500372} (\bibinfo
  {year} {2016})}\BibitemShut {NoStop}%
\bibitem [{\citenamefont {Ishihara}(2010)}]{Ishihara10p011010}%
  \BibitemOpen
  \bibfield  {author} {\bibinfo {author} {\bibfnamefont {S.}~\bibnamefont
  {Ishihara}},\ }\href@noop {} {\bibfield  {journal} {\bibinfo  {journal} {J.
  Phys. Soc. Jpn.}\ }\textbf {\bibinfo {volume} {79}},\ \bibinfo {pages}
  {011010} (\bibinfo {year} {2010})}\BibitemShut {NoStop}%
\bibitem [{\citenamefont {Yamauchi}\ and\ \citenamefont
  {Barone}(2014)}]{Yamauchi14p103201}%
  \BibitemOpen
  \bibfield  {author} {\bibinfo {author} {\bibfnamefont {K.}~\bibnamefont
  {Yamauchi}}\ and\ \bibinfo {author} {\bibfnamefont {P.}~\bibnamefont
  {Barone}},\ }\href@noop {} {\bibfield  {journal} {\bibinfo  {journal} {J.
  Phys. Condens. Matter}\ }\textbf {\bibinfo {volume} {26}},\ \bibinfo {pages}
  {103201} (\bibinfo {year} {2014})}\BibitemShut {NoStop}%
\bibitem [{SM()}]{SM}%
  \BibitemOpen
  \href@noop {} {\enquote {\bibinfo {title} {{S}ee {S}upplemental {M}aterial
  ({SM}) at href for information regarding the computational method, bond
  valence model and parameters, and crystallographic details of the studied
  structures. {SM} cites {R}efs.~\cite{Giannozzi09p395502etalp,
  Liechtenstein95pR5467,Sawada96p12742,Fang03p035101,Biermann05p026404,Czyzyk94p14211,Park17p087602,Monkhorst76p5188,King93p1651,Souza02p117602,Brown09p6858,Brese91p192,Qi16p134308,Xiang09p132408,Verwey39p327,Wexler19p174109}.}}\
  }\BibitemShut {NoStop}%
\bibitem [{\citenamefont {Balachandran}\ and\ \citenamefont
  {Rondinelli}(2013)}]{Balachandran13p054101}%
  \BibitemOpen
  \bibfield  {author} {\bibinfo {author} {\bibfnamefont {P.~V.}\ \bibnamefont
  {Balachandran}}\ and\ \bibinfo {author} {\bibfnamefont {J.~M.}\ \bibnamefont
  {Rondinelli}},\ }\href@noop {} {\bibfield  {journal} {\bibinfo  {journal}
  {Phys. Rev. B}\ }\textbf {\bibinfo {volume} {88}},\ \bibinfo {pages} {054101}
  (\bibinfo {year} {2013})}\BibitemShut {NoStop}%
\bibitem [{\citenamefont {Mazin}\ \emph {et~al.}(2007)\citenamefont {Mazin},
  \citenamefont {Khomskii}, \citenamefont {Lengsdorf}, \citenamefont {Alonso},
  \citenamefont {Marshall}, \citenamefont {Ibberson}, \citenamefont
  {Podlesnyak}, \citenamefont {Mart{\'\i}nez-Lope},\ and\ \citenamefont
  {Abd-Elmeguid}}]{Mazin07p176406}%
  \BibitemOpen
  \bibfield  {author} {\bibinfo {author} {\bibfnamefont {I.~I.}\ \bibnamefont
  {Mazin}}, \bibinfo {author} {\bibfnamefont {D.~I.}\ \bibnamefont {Khomskii}},
  \bibinfo {author} {\bibfnamefont {R.}~\bibnamefont {Lengsdorf}}, \bibinfo
  {author} {\bibfnamefont {J.~A.}\ \bibnamefont {Alonso}}, \bibinfo {author}
  {\bibfnamefont {W.~G.}\ \bibnamefont {Marshall}}, \bibinfo {author}
  {\bibfnamefont {R.~M.}\ \bibnamefont {Ibberson}}, \bibinfo {author}
  {\bibfnamefont {A.}~\bibnamefont {Podlesnyak}}, \bibinfo {author}
  {\bibfnamefont {M.~J.}\ \bibnamefont {Mart{\'\i}nez-Lope}}, \ and\ \bibinfo
  {author} {\bibfnamefont {M.~M.}\ \bibnamefont {Abd-Elmeguid}},\ }\href@noop
  {} {\bibfield  {journal} {\bibinfo  {journal} {Phys. Rev. Lett.}\ }\textbf
  {\bibinfo {volume} {98}},\ \bibinfo {pages} {176406} (\bibinfo {year}
  {2007})}\BibitemShut {NoStop}%
\bibitem [{\citenamefont {Han}\ and\ \citenamefont
  {Millis}(2018)}]{Han18p067601}%
  \BibitemOpen
  \bibfield  {author} {\bibinfo {author} {\bibfnamefont {Q.}~\bibnamefont
  {Han}}\ and\ \bibinfo {author} {\bibfnamefont {A.}~\bibnamefont {Millis}},\
  }\href@noop {} {\bibfield  {journal} {\bibinfo  {journal} {Phys. Rev. Lett.}\
  }\textbf {\bibinfo {volume} {121}},\ \bibinfo {pages} {067601} (\bibinfo
  {year} {2018})}\BibitemShut {NoStop}%
\bibitem [{\citenamefont {Georgescu}\ \emph {et~al.}(2019)\citenamefont
  {Georgescu}, \citenamefont {Peil}, \citenamefont {Disa}, \citenamefont
  {Georges},\ and\ \citenamefont {Millis}}]{Georgescu19p14434}%
  \BibitemOpen
  \bibfield  {author} {\bibinfo {author} {\bibfnamefont {A.~B.}\ \bibnamefont
  {Georgescu}}, \bibinfo {author} {\bibfnamefont {O.~E.}\ \bibnamefont {Peil}},
  \bibinfo {author} {\bibfnamefont {A.~S.}\ \bibnamefont {Disa}}, \bibinfo
  {author} {\bibfnamefont {A.}~\bibnamefont {Georges}}, \ and\ \bibinfo
  {author} {\bibfnamefont {A.~J.}\ \bibnamefont {Millis}},\ }\href@noop {}
  {\bibfield  {journal} {\bibinfo  {journal} {Proc. Natl. Acad. Sci.}\ }\textbf
  {\bibinfo {volume} {116}},\ \bibinfo {pages} {14434} (\bibinfo {year}
  {2019})}\BibitemShut {NoStop}%
\bibitem [{\citenamefont {Georgescu}\ and\ \citenamefont
  {Millis}(2021)}]{Georgescu2021energy}%
  \BibitemOpen
  \bibfield  {author} {\bibinfo {author} {\bibfnamefont {A.~B.}\ \bibnamefont
  {Georgescu}}\ and\ \bibinfo {author} {\bibfnamefont {A.~J.}\ \bibnamefont
  {Millis}},\ }\href@noop {} {\bibfield  {journal} {\bibinfo  {journal} {arXiv
  preprint arXiv:2105.02271}\ } (\bibinfo {year} {2021})}\BibitemShut {NoStop}%
\bibitem [{\citenamefont {Shende}\ \emph {et~al.}(2001)\citenamefont {Shende},
  \citenamefont {Krueger}, \citenamefont {Rossetti},\ and\ \citenamefont
  {Lombardo}}]{Shende01p1648}%
  \BibitemOpen
  \bibfield  {author} {\bibinfo {author} {\bibfnamefont {R.~V.}\ \bibnamefont
  {Shende}}, \bibinfo {author} {\bibfnamefont {D.~S.}\ \bibnamefont {Krueger}},
  \bibinfo {author} {\bibfnamefont {G.~A.}\ \bibnamefont {Rossetti}}, \ and\
  \bibinfo {author} {\bibfnamefont {S.~J.}\ \bibnamefont {Lombardo}},\
  }\href@noop {} {\bibfield  {journal} {\bibinfo  {journal} {J. Am. Ceram.
  Soc.}\ }\textbf {\bibinfo {volume} {84}},\ \bibinfo {pages} {1648} (\bibinfo
  {year} {2001})}\BibitemShut {NoStop}%
\bibitem [{\citenamefont {Ueda}\ \emph {et~al.}(1964)\citenamefont {Ueda},
  \citenamefont {Takiuchi}, \citenamefont {Ikegami},\ and\ \citenamefont
  {Sato}}]{Ueda64p1267}%
  \BibitemOpen
  \bibfield  {author} {\bibinfo {author} {\bibfnamefont {I.}~\bibnamefont
  {Ueda}}, \bibinfo {author} {\bibfnamefont {M.}~\bibnamefont {Takiuchi}},
  \bibinfo {author} {\bibfnamefont {S.}~\bibnamefont {Ikegami}}, \ and\
  \bibinfo {author} {\bibfnamefont {H.}~\bibnamefont {Sato}},\ }\href@noop {}
  {\bibfield  {journal} {\bibinfo  {journal} {J. Phys. Soc. Japan}\ }\textbf
  {\bibinfo {volume} {19}},\ \bibinfo {pages} {1267} (\bibinfo {year}
  {1964})}\BibitemShut {NoStop}%
\bibitem [{\citenamefont {Ikeda}\ \emph {et~al.}(2005)\citenamefont {Ikeda},
  \citenamefont {Ohsumi}, \citenamefont {Ohwada}, \citenamefont {Ishii},
  \citenamefont {Inami}, \citenamefont {Kakurai}, \citenamefont {Murakami},
  \citenamefont {Yoshii}, \citenamefont {Mori}, \citenamefont {Horibe} \emph
  {et~al.}}]{Ikeda05p1136}%
  \BibitemOpen
  \bibfield  {author} {\bibinfo {author} {\bibfnamefont {N.}~\bibnamefont
  {Ikeda}}, \bibinfo {author} {\bibfnamefont {H.}~\bibnamefont {Ohsumi}},
  \bibinfo {author} {\bibfnamefont {K.}~\bibnamefont {Ohwada}}, \bibinfo
  {author} {\bibfnamefont {K.}~\bibnamefont {Ishii}}, \bibinfo {author}
  {\bibfnamefont {T.}~\bibnamefont {Inami}}, \bibinfo {author} {\bibfnamefont
  {K.}~\bibnamefont {Kakurai}}, \bibinfo {author} {\bibfnamefont
  {Y.}~\bibnamefont {Murakami}}, \bibinfo {author} {\bibfnamefont
  {K.}~\bibnamefont {Yoshii}}, \bibinfo {author} {\bibfnamefont
  {S.}~\bibnamefont {Mori}}, \bibinfo {author} {\bibfnamefont {Y.}~\bibnamefont
  {Horibe}},  \emph {et~al.},\ }\href@noop {} {\bibfield  {journal} {\bibinfo
  {journal} {Nature}\ }\textbf {\bibinfo {volume} {436}},\ \bibinfo {pages}
  {1136} (\bibinfo {year} {2005})}\BibitemShut {NoStop}%
\bibitem [{\citenamefont {Ikeda}(2008)}]{Ikeda08p434218}%
  \BibitemOpen
  \bibfield  {author} {\bibinfo {author} {\bibfnamefont {N.}~\bibnamefont
  {Ikeda}},\ }\href@noop {} {\bibfield  {journal} {\bibinfo  {journal} {J.
  Phys. Condens. Matter}\ }\textbf {\bibinfo {volume} {20}},\ \bibinfo {pages}
  {434218} (\bibinfo {year} {2008})}\BibitemShut {NoStop}%
\bibitem [{\citenamefont {Isobe}\ \emph {et~al.}(1990)\citenamefont {Isobe},
  \citenamefont {Kimizuka}, \citenamefont {Iida},\ and\ \citenamefont
  {Takekawa}}]{Isobe90p1917}%
  \BibitemOpen
  \bibfield  {author} {\bibinfo {author} {\bibfnamefont {M.}~\bibnamefont
  {Isobe}}, \bibinfo {author} {\bibfnamefont {N.}~\bibnamefont {Kimizuka}},
  \bibinfo {author} {\bibfnamefont {J.}~\bibnamefont {Iida}}, \ and\ \bibinfo
  {author} {\bibfnamefont {S.}~\bibnamefont {Takekawa}},\ }\href@noop {}
  {\bibfield  {journal} {\bibinfo  {journal} {Acta Crystallogr. Sect. C}\
  }\textbf {\bibinfo {volume} {46}},\ \bibinfo {pages} {1917} (\bibinfo {year}
  {1990})}\BibitemShut {NoStop}%
\bibitem [{\citenamefont {Xiang}\ and\ \citenamefont
  {Whangbo}(2007)}]{Xiang07p246403}%
  \BibitemOpen
  \bibfield  {author} {\bibinfo {author} {\bibfnamefont {H.~J.}\ \bibnamefont
  {Xiang}}\ and\ \bibinfo {author} {\bibfnamefont {M.-H.}\ \bibnamefont
  {Whangbo}},\ }\href@noop {} {\bibfield  {journal} {\bibinfo  {journal} {Phys.
  Rev. Lett.}\ }\textbf {\bibinfo {volume} {98}},\ \bibinfo {pages} {246403}
  (\bibinfo {year} {2007})}\BibitemShut {NoStop}%
\bibitem [{\citenamefont {Xiang}\ \emph {et~al.}(2009)\citenamefont {Xiang},
  \citenamefont {Kan}, \citenamefont {Wei}, \citenamefont {Whangbo},\ and\
  \citenamefont {Yang}}]{Xiang09p132408}%
  \BibitemOpen
  \bibfield  {author} {\bibinfo {author} {\bibfnamefont {H.~J.}\ \bibnamefont
  {Xiang}}, \bibinfo {author} {\bibfnamefont {E.~J.}\ \bibnamefont {Kan}},
  \bibinfo {author} {\bibfnamefont {S.-H.}\ \bibnamefont {Wei}}, \bibinfo
  {author} {\bibfnamefont {M.-H.}\ \bibnamefont {Whangbo}}, \ and\ \bibinfo
  {author} {\bibfnamefont {J.}~\bibnamefont {Yang}},\ }\href@noop {} {\bibfield
   {journal} {\bibinfo  {journal} {Phys. Rev. B}\ }\textbf {\bibinfo {volume}
  {80}},\ \bibinfo {pages} {132408} (\bibinfo {year} {2009})}\BibitemShut
  {NoStop}%
\bibitem [{\citenamefont {de~Groot}\ \emph {et~al.}(2012)\citenamefont
  {de~Groot}, \citenamefont {Mueller}, \citenamefont {Rosenberg}, \citenamefont
  {Keavney}, \citenamefont {Islam}, \citenamefont {Kim},\ and\ \citenamefont
  {Angst}}]{de2012p187601}%
  \BibitemOpen
  \bibfield  {author} {\bibinfo {author} {\bibfnamefont {J.~d.}\ \bibnamefont
  {de~Groot}}, \bibinfo {author} {\bibfnamefont {T.}~\bibnamefont {Mueller}},
  \bibinfo {author} {\bibfnamefont {R.}~\bibnamefont {Rosenberg}}, \bibinfo
  {author} {\bibfnamefont {D.}~\bibnamefont {Keavney}}, \bibinfo {author}
  {\bibfnamefont {Z.}~\bibnamefont {Islam}}, \bibinfo {author} {\bibfnamefont
  {J.-W.}\ \bibnamefont {Kim}}, \ and\ \bibinfo {author} {\bibfnamefont
  {M.}~\bibnamefont {Angst}},\ }\href@noop {} {\bibfield  {journal} {\bibinfo
  {journal} {Phys. Rev. Lett.}\ }\textbf {\bibinfo {volume} {108}},\ \bibinfo
  {pages} {187601} (\bibinfo {year} {2012})}\BibitemShut {NoStop}%
\bibitem [{\citenamefont {Lafuerza}\ \emph {et~al.}(2013)\citenamefont
  {Lafuerza}, \citenamefont {Garc{\'\i}a}, \citenamefont {Sub{\'\i}as},
  \citenamefont {Blasco}, \citenamefont {Conder},\ and\ \citenamefont
  {Pomjakushina}}]{Lafuerza13p085130}%
  \BibitemOpen
  \bibfield  {author} {\bibinfo {author} {\bibfnamefont {S.}~\bibnamefont
  {Lafuerza}}, \bibinfo {author} {\bibfnamefont {J.}~\bibnamefont
  {Garc{\'\i}a}}, \bibinfo {author} {\bibfnamefont {G.}~\bibnamefont
  {Sub{\'\i}as}}, \bibinfo {author} {\bibfnamefont {J.}~\bibnamefont {Blasco}},
  \bibinfo {author} {\bibfnamefont {K.}~\bibnamefont {Conder}}, \ and\ \bibinfo
  {author} {\bibfnamefont {E.}~\bibnamefont {Pomjakushina}},\ }\href@noop {}
  {\bibfield  {journal} {\bibinfo  {journal} {Phys. Rev. B}\ }\textbf {\bibinfo
  {volume} {88}},\ \bibinfo {pages} {085130} (\bibinfo {year}
  {2013})}\BibitemShut {NoStop}%
\bibitem [{\citenamefont {Ruff}\ \emph {et~al.}(2012)\citenamefont {Ruff},
  \citenamefont {Krohns}, \citenamefont {Schrettle}, \citenamefont {Tsurkan},
  \citenamefont {Lunkenheimer},\ and\ \citenamefont {Loidl}}]{Ruff12p290}%
  \BibitemOpen
  \bibfield  {author} {\bibinfo {author} {\bibfnamefont {A.}~\bibnamefont
  {Ruff}}, \bibinfo {author} {\bibfnamefont {S.}~\bibnamefont {Krohns}},
  \bibinfo {author} {\bibfnamefont {F.}~\bibnamefont {Schrettle}}, \bibinfo
  {author} {\bibfnamefont {V.}~\bibnamefont {Tsurkan}}, \bibinfo {author}
  {\bibfnamefont {P.}~\bibnamefont {Lunkenheimer}}, \ and\ \bibinfo {author}
  {\bibfnamefont {A.}~\bibnamefont {Loidl}},\ }\href@noop {} {\bibfield
  {journal} {\bibinfo  {journal} {Eur. Phys. J. B}\ }\textbf {\bibinfo {volume}
  {85}},\ \bibinfo {pages} {290} (\bibinfo {year} {2012})}\BibitemShut
  {NoStop}%
\bibitem [{\citenamefont {Yamauchi}\ and\ \citenamefont
  {Picozzi}(2013)}]{Yamauchi13p113703}%
  \BibitemOpen
  \bibfield  {author} {\bibinfo {author} {\bibfnamefont {K.}~\bibnamefont
  {Yamauchi}}\ and\ \bibinfo {author} {\bibfnamefont {S.}~\bibnamefont
  {Picozzi}},\ }\href@noop {} {\bibfield  {journal} {\bibinfo  {journal} {J.
  Phys. Soc. Jpn.}\ }\textbf {\bibinfo {volume} {82}},\ \bibinfo {pages}
  {113703} (\bibinfo {year} {2013})}\BibitemShut {NoStop}%
\bibitem [{\citenamefont {Senn}\ \emph {et~al.}(2012)\citenamefont {Senn},
  \citenamefont {Wright},\ and\ \citenamefont {Attfield}}]{Senn12p173}%
  \BibitemOpen
  \bibfield  {author} {\bibinfo {author} {\bibfnamefont {M.~S.}\ \bibnamefont
  {Senn}}, \bibinfo {author} {\bibfnamefont {J.~P.}\ \bibnamefont {Wright}}, \
  and\ \bibinfo {author} {\bibfnamefont {J.~P.}\ \bibnamefont {Attfield}},\
  }\href@noop {} {\bibfield  {journal} {\bibinfo  {journal} {Nature}\ }\textbf
  {\bibinfo {volume} {481}},\ \bibinfo {pages} {173} (\bibinfo {year}
  {2012})}\BibitemShut {NoStop}%
\bibitem [{\citenamefont {Verwey}(1939)}]{Verwey39p327}%
  \BibitemOpen
  \bibfield  {author} {\bibinfo {author} {\bibfnamefont {E.~J.~W.}\
  \bibnamefont {Verwey}},\ }\href@noop {} {\bibfield  {journal} {\bibinfo
  {journal} {Nature}\ }\textbf {\bibinfo {volume} {144}},\ \bibinfo {pages}
  {327} (\bibinfo {year} {1939})}\BibitemShut {NoStop}%
\bibitem [{\citenamefont {Ziese}\ \emph {et~al.}(2012)\citenamefont {Ziese},
  \citenamefont {Esquinazi}, \citenamefont {Pantel}, \citenamefont {Alexe},
  \citenamefont {Nemes},\ and\ \citenamefont
  {Garcia-Hern{\'a}ndez}}]{Ziese12p086007}%
  \BibitemOpen
  \bibfield  {author} {\bibinfo {author} {\bibfnamefont {M.}~\bibnamefont
  {Ziese}}, \bibinfo {author} {\bibfnamefont {P.~D.}\ \bibnamefont
  {Esquinazi}}, \bibinfo {author} {\bibfnamefont {D.}~\bibnamefont {Pantel}},
  \bibinfo {author} {\bibfnamefont {M.}~\bibnamefont {Alexe}}, \bibinfo
  {author} {\bibfnamefont {N.~M.}\ \bibnamefont {Nemes}}, \ and\ \bibinfo
  {author} {\bibfnamefont {M.}~\bibnamefont {Garcia-Hern{\'a}ndez}},\
  }\href@noop {} {\bibfield  {journal} {\bibinfo  {journal} {J. Phys. Condens.
  Matter}\ }\textbf {\bibinfo {volume} {24}},\ \bibinfo {pages} {086007}
  (\bibinfo {year} {2012})}\BibitemShut {NoStop}%
\bibitem [{\citenamefont {Giannozzi}\ \emph {et~al.}(2009)\citenamefont
  {Giannozzi}, \citenamefont {Baroni}, \citenamefont {Bonini}, \citenamefont
  {Calandra}, \citenamefont {Car}, \citenamefont {Cavazzoni}, \citenamefont
  {Ceresoli}, \citenamefont {Chiarotti}, \citenamefont {Cococcioni},
  \citenamefont {Dabo}, \citenamefont {Corso}, \citenamefont {de~Gironcoli},
  \citenamefont {Fabris}, \citenamefont {Fratesi}, \citenamefont {Gebauer},
  \citenamefont {Gerstmann}, \citenamefont {Gougoussis}, \citenamefont
  {Kokalj}, \citenamefont {Lazzeri}, \citenamefont {Martin-Samos},
  \citenamefont {Marzari}, \citenamefont {Mauri}, \citenamefont {Mazzarello},
  \citenamefont {Paolini}, \citenamefont {Pasquarello}, \citenamefont
  {Paulatto}, \citenamefont {Sbraccia}, \citenamefont {Scandolo}, \citenamefont
  {Sclauzero}, \citenamefont {Seitsonen}, \citenamefont {Smogunov},
  \citenamefont {Umari},\ and\ \citenamefont
  {Wentzcovitch}}]{Giannozzi09p395502etalp}%
  \BibitemOpen
  \bibfield  {author} {\bibinfo {author} {\bibfnamefont {P.}~\bibnamefont
  {Giannozzi}}, \bibinfo {author} {\bibfnamefont {S.}~\bibnamefont {Baroni}},
  \bibinfo {author} {\bibfnamefont {N.}~\bibnamefont {Bonini}}, \bibinfo
  {author} {\bibfnamefont {M.}~\bibnamefont {Calandra}}, \bibinfo {author}
  {\bibfnamefont {R.}~\bibnamefont {Car}}, \bibinfo {author} {\bibfnamefont
  {C.}~\bibnamefont {Cavazzoni}}, \bibinfo {author} {\bibfnamefont
  {D.}~\bibnamefont {Ceresoli}}, \bibinfo {author} {\bibfnamefont {G.~L.}\
  \bibnamefont {Chiarotti}}, \bibinfo {author} {\bibfnamefont {M.}~\bibnamefont
  {Cococcioni}}, \bibinfo {author} {\bibfnamefont {I.}~\bibnamefont {Dabo}},
  \bibinfo {author} {\bibfnamefont {A.~D.}\ \bibnamefont {Corso}}, \bibinfo
  {author} {\bibfnamefont {S.}~\bibnamefont {de~Gironcoli}}, \bibinfo {author}
  {\bibfnamefont {S.}~\bibnamefont {Fabris}}, \bibinfo {author} {\bibfnamefont
  {G.}~\bibnamefont {Fratesi}}, \bibinfo {author} {\bibfnamefont
  {R.}~\bibnamefont {Gebauer}}, \bibinfo {author} {\bibfnamefont
  {U.}~\bibnamefont {Gerstmann}}, \bibinfo {author} {\bibfnamefont
  {C.}~\bibnamefont {Gougoussis}}, \bibinfo {author} {\bibfnamefont
  {A.}~\bibnamefont {Kokalj}}, \bibinfo {author} {\bibfnamefont
  {M.}~\bibnamefont {Lazzeri}}, \bibinfo {author} {\bibfnamefont
  {L.}~\bibnamefont {Martin-Samos}}, \bibinfo {author} {\bibfnamefont
  {N.}~\bibnamefont {Marzari}}, \bibinfo {author} {\bibfnamefont
  {F.}~\bibnamefont {Mauri}}, \bibinfo {author} {\bibfnamefont
  {R.}~\bibnamefont {Mazzarello}}, \bibinfo {author} {\bibfnamefont
  {S.}~\bibnamefont {Paolini}}, \bibinfo {author} {\bibfnamefont
  {A.}~\bibnamefont {Pasquarello}}, \bibinfo {author} {\bibfnamefont
  {L.}~\bibnamefont {Paulatto}}, \bibinfo {author} {\bibfnamefont
  {C.}~\bibnamefont {Sbraccia}}, \bibinfo {author} {\bibfnamefont
  {S.}~\bibnamefont {Scandolo}}, \bibinfo {author} {\bibfnamefont
  {G.}~\bibnamefont {Sclauzero}}, \bibinfo {author} {\bibfnamefont {A.~P.}\
  \bibnamefont {Seitsonen}}, \bibinfo {author} {\bibfnamefont {A.}~\bibnamefont
  {Smogunov}}, \bibinfo {author} {\bibfnamefont {P.}~\bibnamefont {Umari}}, \
  and\ \bibinfo {author} {\bibfnamefont {R.~M.}\ \bibnamefont {Wentzcovitch}},\
  }\href@noop {} {\bibfield  {journal} {\bibinfo  {journal} {J. Phys.: Condens.
  Matter}\ }\textbf {\bibinfo {volume} {21}},\ \bibinfo {pages} {395502}
  (\bibinfo {year} {2009})}\BibitemShut {NoStop}%
\bibitem [{\citenamefont {Liechtenstein}\ \emph {et~al.}(1995)\citenamefont
  {Liechtenstein}, \citenamefont {Anisimov},\ and\ \citenamefont
  {Zaanen}}]{Liechtenstein95pR5467}%
  \BibitemOpen
  \bibfield  {author} {\bibinfo {author} {\bibfnamefont {A.}~\bibnamefont
  {Liechtenstein}}, \bibinfo {author} {\bibfnamefont {V.~I.}\ \bibnamefont
  {Anisimov}}, \ and\ \bibinfo {author} {\bibfnamefont {J.}~\bibnamefont
  {Zaanen}},\ }\href@noop {} {\bibfield  {journal} {\bibinfo  {journal} {Phys.
  Rev. B}\ }\textbf {\bibinfo {volume} {52}},\ \bibinfo {pages} {R5467}
  (\bibinfo {year} {1995})}\BibitemShut {NoStop}%
\bibitem [{\citenamefont {Sawada}\ \emph {et~al.}(1996)\citenamefont {Sawada},
  \citenamefont {Hamada}, \citenamefont {Terakura},\ and\ \citenamefont
  {Asada}}]{Sawada96p12742}%
  \BibitemOpen
  \bibfield  {author} {\bibinfo {author} {\bibfnamefont {H.}~\bibnamefont
  {Sawada}}, \bibinfo {author} {\bibfnamefont {N.}~\bibnamefont {Hamada}},
  \bibinfo {author} {\bibfnamefont {K.}~\bibnamefont {Terakura}}, \ and\
  \bibinfo {author} {\bibfnamefont {T.}~\bibnamefont {Asada}},\ }\href@noop {}
  {\bibfield  {journal} {\bibinfo  {journal} {Phys. Rev. B}\ }\textbf {\bibinfo
  {volume} {53}},\ \bibinfo {pages} {12742} (\bibinfo {year}
  {1996})}\BibitemShut {NoStop}%
\bibitem [{\citenamefont {Fang}\ \emph {et~al.}(2003)\citenamefont {Fang},
  \citenamefont {Nagaosa},\ and\ \citenamefont {Terakura}}]{Fang03p035101}%
  \BibitemOpen
  \bibfield  {author} {\bibinfo {author} {\bibfnamefont {Z.}~\bibnamefont
  {Fang}}, \bibinfo {author} {\bibfnamefont {N.}~\bibnamefont {Nagaosa}}, \
  and\ \bibinfo {author} {\bibfnamefont {K.}~\bibnamefont {Terakura}},\
  }\href@noop {} {\bibfield  {journal} {\bibinfo  {journal} {Phys. Rev. B}\
  }\textbf {\bibinfo {volume} {67}},\ \bibinfo {pages} {035101} (\bibinfo
  {year} {2003})}\BibitemShut {NoStop}%
\bibitem [{\citenamefont {Biermann}\ \emph {et~al.}(2005)\citenamefont
  {Biermann}, \citenamefont {Poteryaev}, \citenamefont {Lichtenstein},\ and\
  \citenamefont {Georges}}]{Biermann05p026404}%
  \BibitemOpen
  \bibfield  {author} {\bibinfo {author} {\bibfnamefont {S.}~\bibnamefont
  {Biermann}}, \bibinfo {author} {\bibfnamefont {A.}~\bibnamefont {Poteryaev}},
  \bibinfo {author} {\bibfnamefont {A.}~\bibnamefont {Lichtenstein}}, \ and\
  \bibinfo {author} {\bibfnamefont {A.}~\bibnamefont {Georges}},\ }\href@noop
  {} {\bibfield  {journal} {\bibinfo  {journal} {Phys. Rev. Lett.}\ }\textbf
  {\bibinfo {volume} {94}},\ \bibinfo {pages} {026404} (\bibinfo {year}
  {2005})}\BibitemShut {NoStop}%
\bibitem [{\citenamefont {Czy{\.z}yk}\ and\ \citenamefont
  {Sawatzky}(1994)}]{Czyzyk94p14211}%
  \BibitemOpen
  \bibfield  {author} {\bibinfo {author} {\bibfnamefont {M.}~\bibnamefont
  {Czy{\.z}yk}}\ and\ \bibinfo {author} {\bibfnamefont {G.}~\bibnamefont
  {Sawatzky}},\ }\href@noop {} {\bibfield  {journal} {\bibinfo  {journal}
  {Phys. Rev. B}\ }\textbf {\bibinfo {volume} {49}},\ \bibinfo {pages} {14211}
  (\bibinfo {year} {1994})}\BibitemShut {NoStop}%
\bibitem [{\citenamefont {Monkhorst}\ and\ \citenamefont
  {Pack}(1976)}]{Monkhorst76p5188}%
  \BibitemOpen
  \bibfield  {author} {\bibinfo {author} {\bibfnamefont {H.~J.}\ \bibnamefont
  {Monkhorst}}\ and\ \bibinfo {author} {\bibfnamefont {J.~D.}\ \bibnamefont
  {Pack}},\ }\href@noop {} {\bibfield  {journal} {\bibinfo  {journal} {Phys.
  Rev. B}\ }\textbf {\bibinfo {volume} {13}},\ \bibinfo {pages} {5188}
  (\bibinfo {year} {1976})}\BibitemShut {NoStop}%
\bibitem [{\citenamefont {King-Smith}\ and\ \citenamefont
  {Vanderbilt}(1993)}]{King93p1651}%
  \BibitemOpen
  \bibfield  {author} {\bibinfo {author} {\bibfnamefont {R.~D.}\ \bibnamefont
  {King-Smith}}\ and\ \bibinfo {author} {\bibfnamefont {D.}~\bibnamefont
  {Vanderbilt}},\ }\href@noop {} {\bibfield  {journal} {\bibinfo  {journal}
  {Phys. Rev. B}\ }\textbf {\bibinfo {volume} {47}},\ \bibinfo {pages} {1651}
  (\bibinfo {year} {1993})}\BibitemShut {NoStop}%
\bibitem [{\citenamefont {Souza}\ \emph {et~al.}(2002)\citenamefont {Souza},
  \citenamefont {{\'I}niguez},\ and\ \citenamefont
  {Vanderbilt}}]{Souza02p117602}%
  \BibitemOpen
  \bibfield  {author} {\bibinfo {author} {\bibfnamefont {I.}~\bibnamefont
  {Souza}}, \bibinfo {author} {\bibfnamefont {J.}~\bibnamefont {{\'I}niguez}},
  \ and\ \bibinfo {author} {\bibfnamefont {D.}~\bibnamefont {Vanderbilt}},\
  }\href@noop {} {\bibfield  {journal} {\bibinfo  {journal} {Phys. Rev. Lett.}\
  }\textbf {\bibinfo {volume} {89}},\ \bibinfo {pages} {117602} (\bibinfo
  {year} {2002})}\BibitemShut {NoStop}%
\bibitem [{\citenamefont {Brown}(2009)}]{Brown09p6858}%
  \BibitemOpen
  \bibfield  {author} {\bibinfo {author} {\bibfnamefont {I.~D.}\ \bibnamefont
  {Brown}},\ }\href@noop {} {\bibfield  {journal} {\bibinfo  {journal} {Chem.
  Rev.}\ }\textbf {\bibinfo {volume} {109}},\ \bibinfo {pages} {6858} (\bibinfo
  {year} {2009})}\BibitemShut {NoStop}%
\bibitem [{\citenamefont {Brese}\ and\ \citenamefont
  {O'keeffe}(1991)}]{Brese91p192}%
  \BibitemOpen
  \bibfield  {author} {\bibinfo {author} {\bibfnamefont {N.}~\bibnamefont
  {Brese}}\ and\ \bibinfo {author} {\bibfnamefont {M.}~\bibnamefont
  {O'keeffe}},\ }\href@noop {} {\bibfield  {journal} {\bibinfo  {journal} {Acta
  Crystallogr. B Struct. Sci.}\ }\textbf {\bibinfo {volume} {47}},\ \bibinfo
  {pages} {192} (\bibinfo {year} {1991})}\BibitemShut {NoStop}%
\bibitem [{\citenamefont {Qi}\ \emph {et~al.}(2016)\citenamefont {Qi},
  \citenamefont {Liu}, \citenamefont {Grinberg},\ and\ \citenamefont
  {Rappe}}]{Qi16p134308}%
  \BibitemOpen
  \bibfield  {author} {\bibinfo {author} {\bibfnamefont {Y.}~\bibnamefont
  {Qi}}, \bibinfo {author} {\bibfnamefont {S.}~\bibnamefont {Liu}}, \bibinfo
  {author} {\bibfnamefont {I.}~\bibnamefont {Grinberg}}, \ and\ \bibinfo
  {author} {\bibfnamefont {A.~M.}\ \bibnamefont {Rappe}},\ }\href@noop {}
  {\bibfield  {journal} {\bibinfo  {journal} {Phys. Rev. B}\ }\textbf {\bibinfo
  {volume} {94}},\ \bibinfo {pages} {134308} (\bibinfo {year}
  {2016})}\BibitemShut {NoStop}%
\bibitem [{\citenamefont {Wexler}\ \emph {et~al.}(2019)\citenamefont {Wexler},
  \citenamefont {Qi},\ and\ \citenamefont {Rappe}}]{Wexler19p174109}%
  \BibitemOpen
  \bibfield  {author} {\bibinfo {author} {\bibfnamefont {R.~B.}\ \bibnamefont
  {Wexler}}, \bibinfo {author} {\bibfnamefont {Y.}~\bibnamefont {Qi}}, \ and\
  \bibinfo {author} {\bibfnamefont {A.~M.}\ \bibnamefont {Rappe}},\ }\href@noop
  {} {\bibfield  {journal} {\bibinfo  {journal} {Phys. Rev. B}\ }\textbf
  {\bibinfo {volume} {100}},\ \bibinfo {pages} {174109} (\bibinfo {year}
  {2019})}\BibitemShut {NoStop}%
\end{thebibliography}%

\end{document}